\documentclass[%
aps
prb
amsmath,amssymb,
reprint,%
]{revtex4-1}

\usepackage{graphicx}
\usepackage{dcolumn}
\usepackage{bm}
\usepackage[section]{placeins}
\usepackage{color}
\usepackage{mathrsfs}
\usepackage{ulem}
\usepackage{hyperref}

\newcommand{\tr}[1]{\textcolor{black}{#1}}
\newcommand{\trr}[1]{\textcolor{black}{#1}}
\newcommand{\tb}[1]{\textcolor{black}{#1}}
\newcommand{\tg}[1]{\textcolor{black}{#1}}

\begin{document}


\title{Enhancement of superconductivity on thin film of Sn under high pressure}
\author{Misaki Sasaki$^{1}$}
\author{Masahiro Ohkuma$^{2}$}
\altaffiliation[Present address:~]{School of Engineering, \tr{Institute of Science Tokyo}, Yokohama, Kanagawa 226-8501, Japan}
\author{Ryo Matsumoto$^{2}$}
\author{Toru Shinmei$^{3}$}
\author{Tetsuo Irifune$^{3}$}
\author{Yoshihiko Takano$^{2}$}
\author{Katsuya Shimizu$^{1}$}
\affiliation{$^{1}$KYOKUGEN, Graduate School of Engineering Science, Osaka University, Toyonaka, Osaka, 560-8531, Japan}
\affiliation{$^{2}$Research Center for Materials Nanoarchitectonics (MANA), National Institute for Materials Science, Tsukuba 305-0047, Ibaraki, Japan}
\affiliation{$^{3}$Geodynamics Research Center, Ehime University, Matsuyama, 790-8577, Ehime, Japan}

\date{\today}

\begin{abstract}
We investigated \tg{the} pressure effects of a superconductivity on thin films of Sn.
\tg{Elemental} superconductor Sn with \tg{a} body-centered tetragonal structure, $\beta$-Sn, exhibits superconductivity below the superconducting transition temperature ($T_{\rm c}=3.72$ K) at ambient pressure.
$T_{\rm c}$ of Sn increases with lowering dimension such as \tg{in} thin film and \tg{nanowire growth}, or by \tg{high-pressure application}.
\tg{For thin films}, $T_{\rm c}$ \tg{exhibits a} slight increase up to approximately 4 K \tg{compared to} the bulk value, \tg{attributable} to the crystalline size and lattice disorder.
By applying pressure on a bulk Sn, \trr{$T_{\rm c}$ initially decreases from 3.72 K as the pressure increases.
Further increasing pressure up to 10 GPa, $T_{\rm c}$ increases} to 5.3 K with the structural transformation.
However, the combination \tg{of these effects} on \tg{thin films of Sn, namely, thin-film growth and pressure effects, remains underexplored}.
In this study, we combined \tg{film-growth} and \tb{pressure-application} techniques \tg{to} further \tb{increase} $T_{\rm c}$ using \tb{a} diamond anvil cell with \tb{boron-doped} diamond electrodes.
The drop of the electrical resistance suggesting the onset of $T_{\rm c}$ on the thin film reached above 6 K in $\gamma$-Sn phase.
\tb{Further}, the upper critical magnetic field was drastically enhanced.
\tb{Atomic} force microscopy suggests that the refinement of the grain size of the thin film \tb{under} the non-hydrostatic pressure \tb{conditions} contributes to \tb{stabilizing} the higher $T_{\rm c}$ of $\gamma$-Sn.
\end{abstract}
%
%
\maketitle
\section{Introduction}
{Applying high \tb{pressures} to directly compress a material is a useful approach to investigate intriguing physical properties and search for new materials \cite{maoSolidsLiquidsGases2018, yamanakaSiliconClathratesCarbon2010}.
For instance, oxygen\tb{---a gas at ambient conditions---} \tb{exhibits} metallic behaviors and superconductivity \tb{at high pressures} \cite{shimizuSuperconductivityOxygen1998a, shimizuPressureinducedSuperconductivityElemental2005}.
Recently, \tb{high-temperature} superconductors such as \tb{hydrogen-rich} materials and nickelates under high pressure have attracted \tb{considerable} attention \cite{drozdovConventionalSuperconductivity2032015, drozdovSuperconductivity250Lanthanum2019, somayazuluEvidenceSuperconductivity2602019, matsumotoElectricalTransportMeasurements2020, semenokSuperconductivity161Thorium2020, chenHighTemperatureSuperconductingPhases2021, kongSuperconductivity243Yttriumhydrogen2021, troyanAnomalousHighTemperatureSuperconductivity2021, liSuperconductivity200Discovered2022, maHighTemperatureSuperconductingPhase2022, songStoichiometricTernarySuperhydride2023, crossHightemperatureSuperconductivity$mathrmLa_4mathrmH_23$2024, sunSignaturesSuperconductivity802023, sakakibaraTheoreticalAnalysisPossibility2024}.
In addition, recent discoveries of the \tb{high-temperature} superconducting \tb{states in} elemental solid Ti and Sc \tb{at} extreme \tb{pressures} imply \tb{the} potential of \tb{high-temperature} superconductivity in \tb{high-pressure} phases of other elements \cite{liu$T_c$236Robust2022, zhangRecordHighTc2022, yingRecordHigh362023}.}

Some elemental superconductors with \tb{thin-film dimensions} show \tb{an increase in} superconducting transition temperature ($T_{\rm c}$) \tb{compared to the} bulk value, \tb{attributable to the} crystalline size and lattice disorder \cite{buckelEinflussKondensationBei1954a, abelesEnhancementSuperconductivityMetal1966, garlandEffectLatticeDisorder1968, stronginEnhancedSuperconductivityLayered1968}.
In the case of Sn, $T_{\rm c}$ \tb{for} thin films and {nano-wires} varies slightly depending on \tb{the size and surface morphology} \cite{tianDissipationQuasionedimensionalSuperconducting2005, beutelMicrowaveStudySuperconducting2016, bangCharacterizationSuperconductingSn2019, lozanoExperimentalObservationElectronphonon2019a}.
The mechanism \tb{of $T_{\rm c}$ enhancement} is not thoroughly clear \tb{although it has been} proposed to \tb{arise} from changes in the phonon density of states, the electron density of states, and \tb{the} {electron-phonon coupling} \cite{knorrSuperconductivityPhononSpectra1970, houbenLatticeDynamicsSn2017a, houbenInfluencePhononSoftening2020a}.

Sn exhibits various {crystal structures} \tb{at high pressures} \cite{barnettXRayDiffractionStudies2004, salamatHighpressureStructuralTransformations2013, deffrennesTinSnHigh2022}.
\tb{Half a century ago,} Wittig investigated \tb{the} electrical transport properties \tb{of} bulk Sn \tb{at high pressures of} up to 16 GPa and revealed the highest $T_{\rm c}$ is 5.3 K at 11.3 GPa on the pressure induced phase, $\gamma$-Sn, where $\beta$-Sn shows  superconductivity below 3.72 K at ambient pressure \cite{wittigSupraleitungZinnUnd1966}.
However, \tb{research on} combination \tb{of} pressure \tb{application} and thin-film \tb{growth} on elemental superconductors \tb{is inadequate}.
Here, we hypothesize that changing the crystalline size \tb{of a thin film via pressure application} could stabilize the higher $T_{\rm c}$.

In this \tb{study}, we combined \tb{thin} film growth and \tb{pressure-application} techniques \tb{to increase} $T_{\rm c}$ using a diamond anvil cell (DAC) with \tb{boron-doped} diamond (BDD) electrodes \cite{matsumotoNoteNovelDiamond2016, matsumotoDiamondAnvilCell2017, matsumotoDiamondAnvilCells2018}.
We investigated the pressure effects \tb{on} superconductivity of thin films of Sn compared \tb{to} the bulk sample.
We observed \tb{a} higher $T_{\rm c}$ \tb{for} the thin film \tb{comapred to} previous \tb{studies on high-pressure} phase.
\tb{Further}, we observed \tb{a} drastically enhanced critical magnetic field on the thin film under high pressure.

\section{Experimental procedure}
Thin films of Sn were deposited on a diamond anvil by a resistance heating evaporation.
\tb{The} target metal was a \tb{high-purity} Sn, purchased from Kojundo Chemical Lab. Co. Ltd.
Comparing with the high pressure measurements, we also prepared \tb{Sn} thin film on \tb{a} diamond substrate for electrical transport measurements at ambient pressure.
The film deposition on two diamonds was performed simultaneously.
{The optical images of the thin films on the diamond anvil and the diamond substrate are shown in Fig.~\ref{f1}(a) and the inset of Fig.~\ref{f1}(c)\tb{, respectively}}.
The \tb{film thickness} and surface morphology of the films were evaluated \tb{via} atomic force \tb{microscopy} (\tb{AFM;} Nanocute, SII NanoTechnology Inc.) at room temperature.
For \tb{the} magnetic measurements \tb{of the} bulk Sn, a wire of \tb{high-purity} Sn (Nilaco Corp.) was used.

\tb{High-pressure} generation was performed using DAC.
The pressure value at room temperature was evaluated using ruby fluorescence and Raman shift of diamond \cite{maoSpecificVolumeMeasurements2008, akahamaHighpressureRamanSpectroscopy2004}.
In the electrical transport measurements, diamond anvil with BDD electrodes \tb{with a} culet size of 300 $\mu$m was used \cite{matsumotoNoteNovelDiamond2016, matsumotoDiamondAnvilCell2017, matsumotoDiamondAnvilCells2018}.
\tr{The BDD electrodes were deposited homoepitaxially on the diamond anvil by microwave plasma chemical vapor deposition \cite{takanoSuperconductivityDiamondThin2004}.
This electrode exhibits high durability and can be reused until the diamond anvil itself fractures.
Further, thin films could be deposited directly on the diamond anvil with BDD electrodes, eliminating the need for an electrode fabrication process after thin film deposition \cite{matsumotoNoteNovelDiamond2016,adachiDemonstrationElectricDouble2020,  matsumotoSynthesisElectricalTransport2021}.
The pressure-transmitting media in the solid, liquid, and gaseous states are compatible with this system.}
A gasket of a stainless steel was pre-indented and \tb{a} 200 $\mu$m diameter hole \tb{was drilled}.
The insulating layer was prepared using \tb{a} \tb{MgO--epoxy} mixture.
We termed this setup as non-hydrostatic.
We also performed the \tb{high-pressure} generation with better hydrostatic pressure condition \tb{than non-hydrostatic measurement} using \tb{a} liquid \tb{pressure-transmitting} medium, glycerol.
\tb{A 150 $\mu$m diameter hole was drilled in the insulating layer of MgO--epoxy mixture to prepare the sample space, which was filled with glycerol.
We termed this setup as quasi-hydrostatic.}
In quasi-hydrostatic pressure measurements, the pressure value was evaluated using Raman shift of diamond \cite{akahamaHighpressureRamanSpectroscopy2004}.
The electrical transports under a magnetic field ($H$) perpendicular to the surface were measured by a four-terminal method by physical properties measurement system (Quantum Design).

\tb{For} magnetic measurements, we used \tb{a} miniature DAC in combination with a superconducting quantum interference device magnetometer {(MPMS, Quantum Design)} \cite{mitoDevelopmentMiniatureDiamond2001, mitoEffectiveDisappearanceMeissner2014, mitoUniaxialStrainEffects2016, mitoUniaxialStrainEffects2017, abdel-hafiezHighpressureEffectsIsotropic2018, mitoHydrostaticPressureEffects2019}.
A nano-polycrystalline diamond with culet size of 600 $\mu$m and a pre-indented tungsten gasket with a hole size of 200 $\mu$m were used \cite{irifuneUltrahardPolycrystallineDiamond2003}.
\tb{Bulk Sn pieces} and ruby powders were loaded into the sample space without \tb{a} pressure-transmitting medium.
The in-phase component of \tb{the AC} magnetic response ($m'$) was measured.
The frequency and amplitude of the \tb{AC} field were 3 Hz and 0.2 mT, respectively.

\section{Experimental Results}
\subsection{Ambient pressure}

\begin{figure}[htb!]
\centering\includegraphics[clip,width=1\columnwidth]{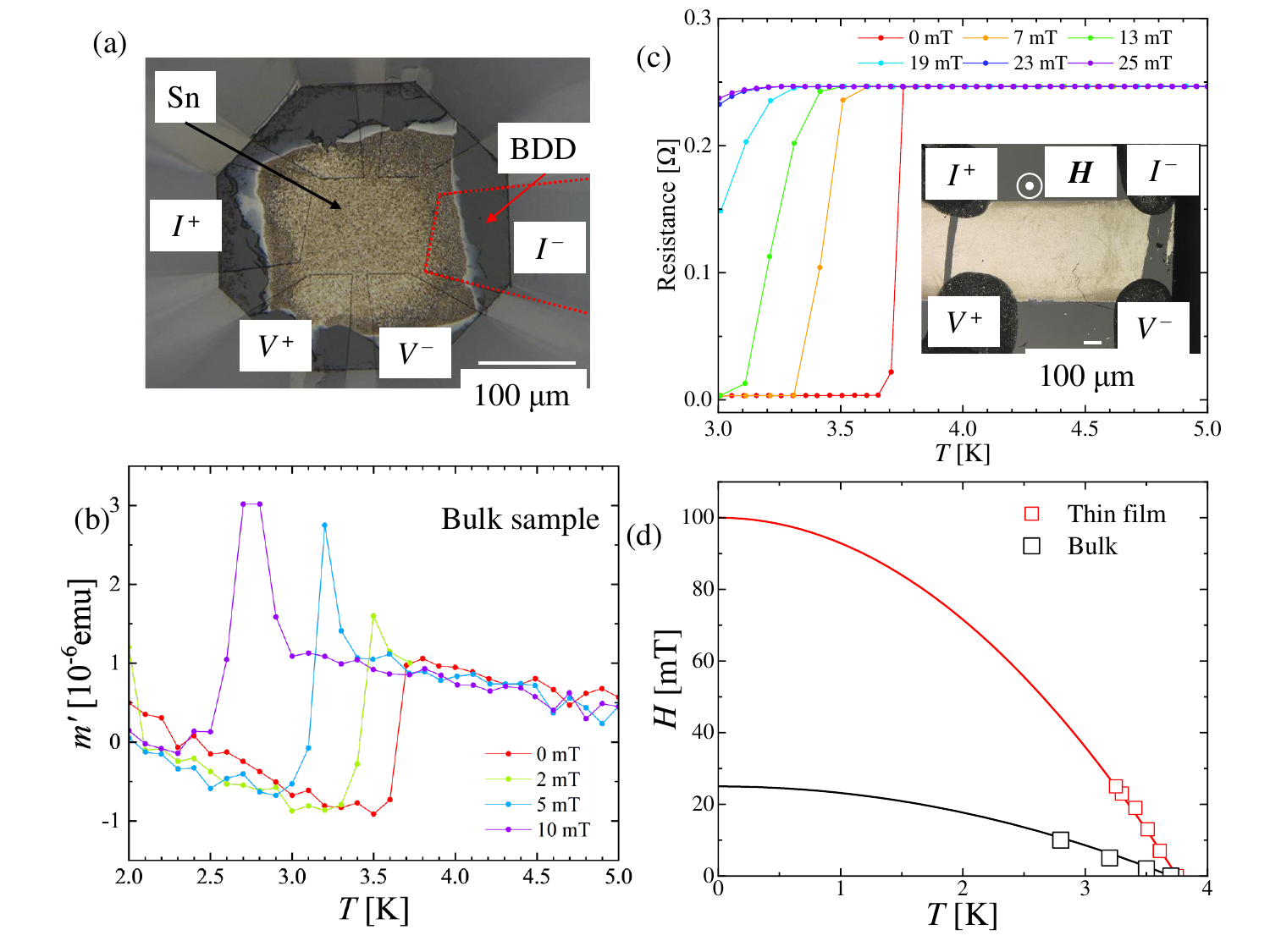}
\caption{\label{f1} (Color online) (a) \tb{Optical} image of {the thin film of Sn} on the diamond anvil with BDD electrodes.
{The dotted area indicates one of the BDD electrodes.}
(b) $T$ dependence of \tb{the} in-phase component of $m'$ on the bulk Sn.
(c) $T$ dependence of $R$ on {the thin film of Sn} on the diamond substrate.
The magnetic field was applied perpendicularly to film.
The inset shows the optical image of {the thin film of Sn} on the diamond substrate
\trr{The thickness of the thin film was approximately 100 nm.}
(d) $T$ dependence of \tb{the} critical magnetic field on thin film and bulk Sn.}
\end{figure}

Figure~\ref{f1}(b) shows the temperature ($T$) dependence of $m'$ \tb{for} the bulk Sn under $H$, \tr{where no background signal from DAC is subtracted.}
Below 3.7 K, the diamagnetic signal suggesting the superconducting state was observed at $H=0$.
The \tb{$T_{\rm c}$ onset} was decreased by applying $H$.
Figure~\ref{f1}(c) shows \tb{the} $T$ dependence of the electrical resistance ($R$) under $H$ \tb{perpendicular} to the film \tb{surface}.
The residual resistance ratio (RRR) was {estimated to be 11.}
At $H=0$, \tb{an $R$ drop} suggesting the onset of the superconducting transition was observed at 3.75 K.
$T_{\rm c}$ slightly increased \tb{compared to} the bulk value and was similar to \tb{values from} the previous studies on thin films \cite{beutelMicrowaveStudySuperconducting2016, bangCharacterizationSuperconductingSn2019}.
By applying $H$, the onset of $T_{\rm c}$ decreased with increasing $H$.
However, the critical magnetic field was three times higher than that of the bulk \tb{Sn} \tb{(Fig.~\ref{f1}(d))}.
{$H_{\rm c}$ of 100 mT estimated \tb{using} $H_{\rm c}(T)=H_{\rm c}(1-(T/T_{\rm c})^2)$ was similar to \tb{that of} a previous study \cite{bangCharacterizationSuperconductingSn2019}.}
Considering to the $H_{\rm c}$, the thin film transformed \tb{into a} type II superconductor, as previously reported \cite{dolanCriticalThicknessesSuperconducting1973, bangCharacterizationSuperconductingSn2019, ohkumaNonreciprocalSupercurrentThin2023}.

\subsection{Non-hydrostatic pressure on bulk Sn}
\begin{figure}[htb!]
\centering\includegraphics[clip,width=0.7\columnwidth]{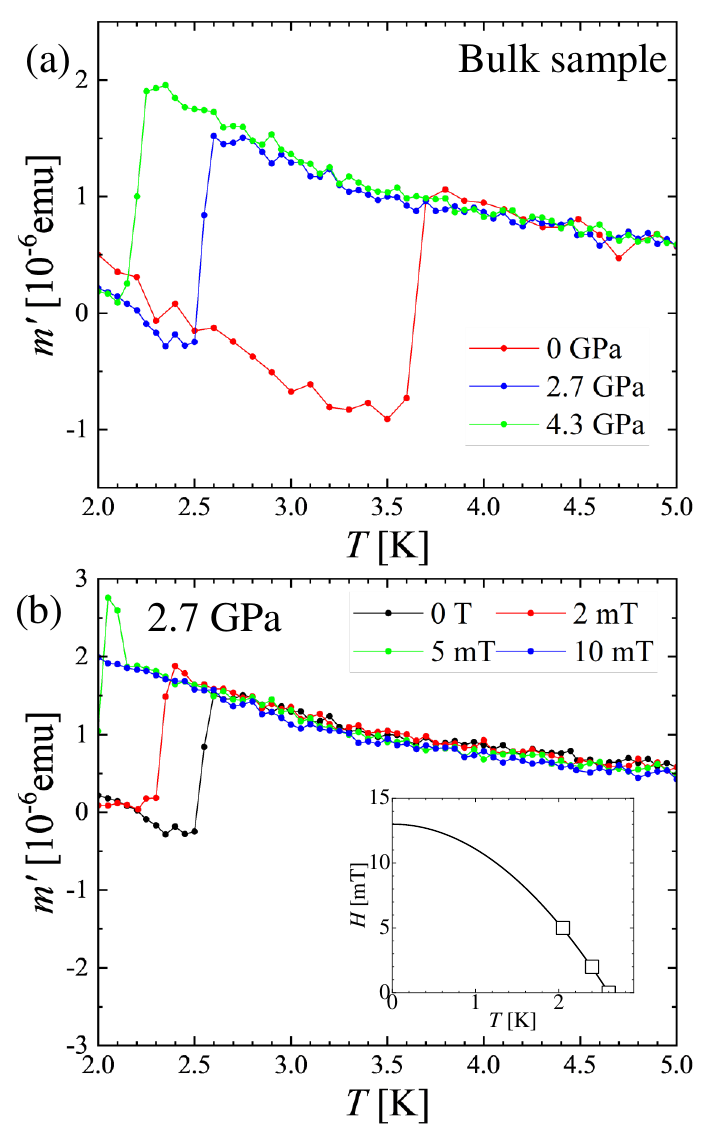}
\caption{\label{f_m} (Color online) (a) $T$ dependence of $m'$ \tb{on bulk Sn under high pressures}.
(b) $T$ dependence of $m'$ at 2.7 GPa under \tb{varying} magnetic fields.
The inset shows \tb{the} $T$ dependence of the critical magnetic field.
}
\end{figure}

Figure~\ref{f_m}(a) shows \tb{the} $T$ dependence of $m'$ on the bulk Sn under high \tb{pressures}.
\tb{On} applying pressure, $T_{\rm c}$ decreased, as previously reported.
Figure~\ref{f_m}(b) shows \tb{the} $T$ dependence of $m'$ at 2.7 GPa under \tb{varying} $H$.
The decrease of $T_{\rm c}$ was observed by applying $H=2$ mT.
At $H=10$ mT, $T_{\rm c}$ was below 2 K.
$H_{\rm c}$ was evaluated to be 13 mT, as shown in the inset of Fig.~\ref{f_m}(b).

\subsection{Non-hydrostatic pressure on thin film}
\begin{figure}[htb!]
\centering\includegraphics[clip,width=0.7\columnwidth]{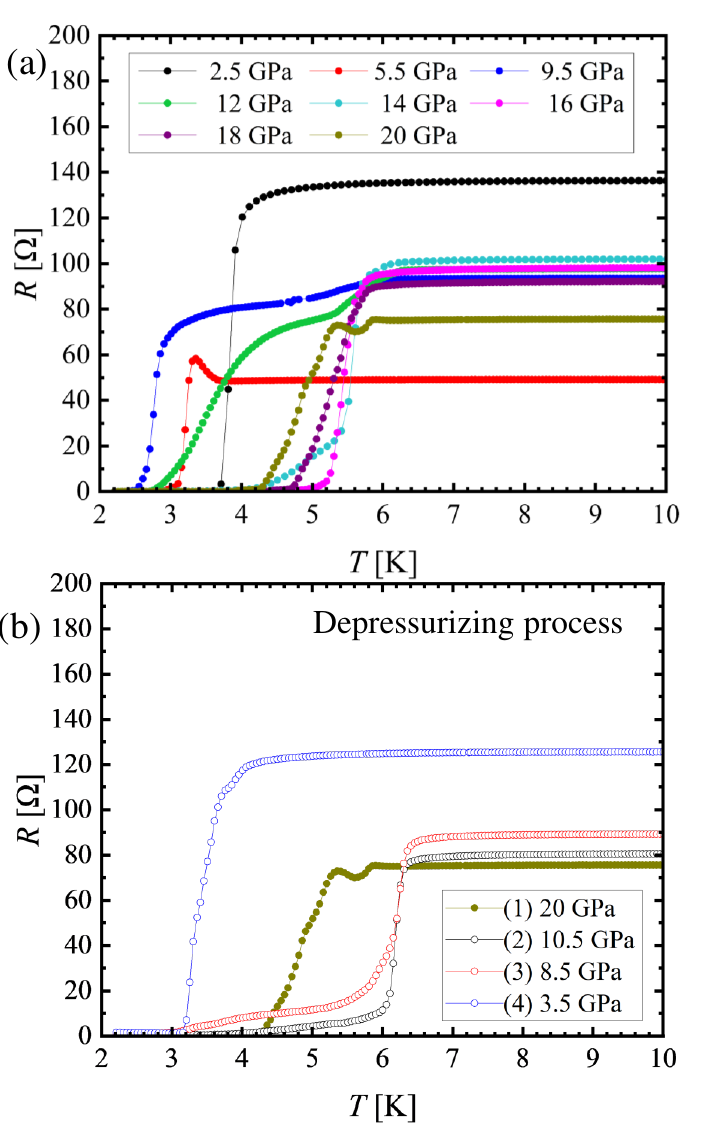}
\caption{\label{f_RT_P} (Color online) $T$ dependence of electrical $R$ under {non-hydrostatic pressure} with (a) pressurizing process (b) depressurizing process.
\tr{The number in (b) indicates the order of measurements.}}
\end{figure}

\begin{figure}[htb!]
\centering\includegraphics[clip,width=0.7\columnwidth]{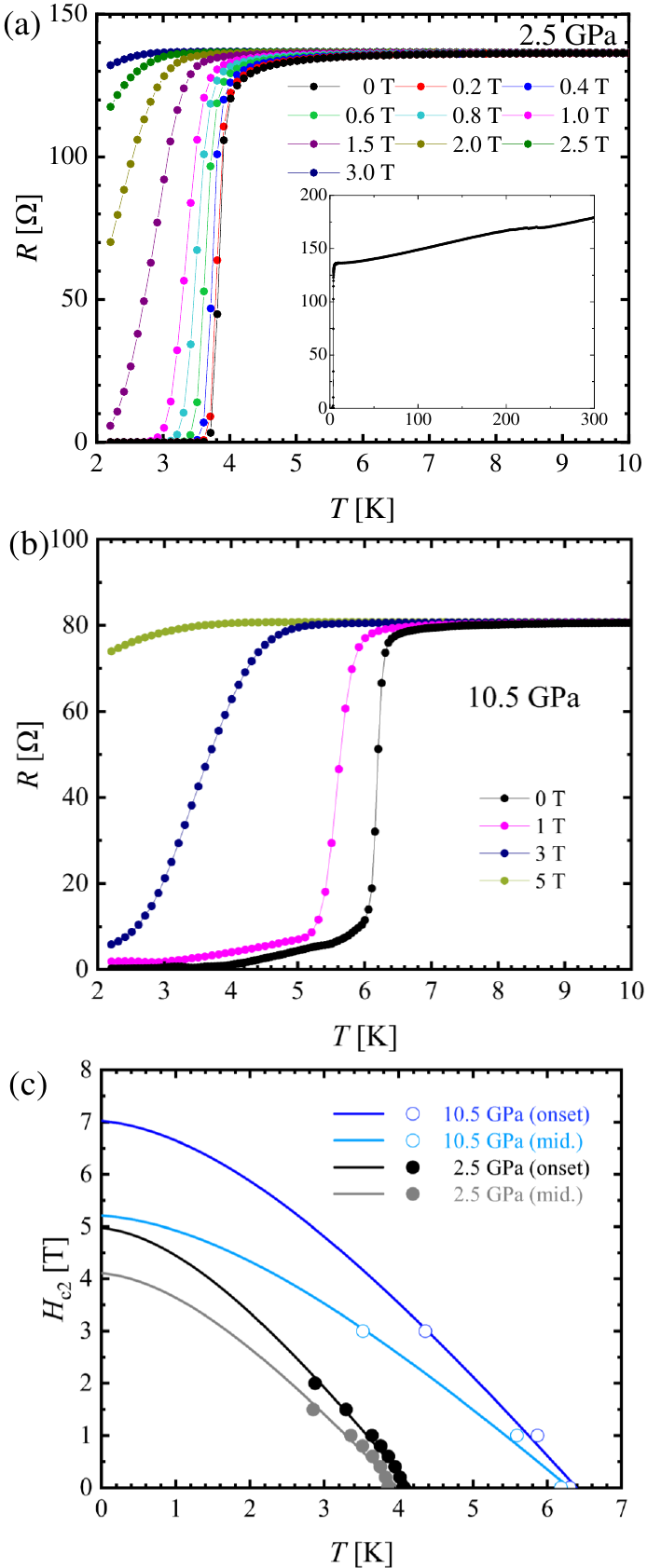}
\caption{\label{f_RT_H} (Color online) (a) and (b) $T$ dependence of $R$ under \tb{the} magnetic field at (a) 2.5 and (b) 10.5 GPa.
{The inset of (a) shows $T$ dependence of $R$ between 2 and 300 K at 2.5 GPa without the external magnetic field.}
(c) $T$ dependence of the upper critical magnetic field at 2.5 and 10.5 GPa.
The solid lines represent the fitting curves estimated by \tb{the} WHH model.
}
\end{figure}

Figure~\ref{f_RT_P}(a) shows \tb{the} $T$ dependence of $R$ on the {thin film of Sn} under high pressures \tb{on pressurization}.
{As shown in the inset of Fig.~\ref{f_RT_H}(a), RRR was estimated to be 1.4 at 2.5 GPa, which was much lower than that \tb{under} ambient pressure.}
We speculate that the crystallographic defect was introduced \tb{because of} the non-hydrostatic pressure \tb{condition}.
The $R$ drop suggesting the superconducting transition was observed around 4 K, whereas $T_{\rm c}$ was 3.75 K under ambient pressure.
$T_{\rm c}$ decreased \tb{on} further increasing \tb{the} pressure, as previously reported.
$R$ showed peak behavior just above $T_{\rm c}$ at 5.5 GPa, possibly \tb{due} to the granularity or disorder on the thin films.
Above 9.5 GPa, $R$ slightly tended to decrease around 6 K, suggesting the superconducting transition, where $\gamma$-Sn phase could emerge.
\tb{With further pressure application,} the \tb{$R$ drop} became significant, and $T_{\rm c}$ slightly decreased.
After applying 20 GPa, the pressure was decreased to 10.5 GPa.
Figure~\ref{f_RT_P}(b) shows \tb{the} $T$ dependence of $R$ for the {thin film of Sn} under high pressures with \tb{depressurization}.
The onset of $T_{\rm c}$ increased to {6.3 K} at 10.5 GPa, and $\gamma$-Sn remained at 8.5 GPa.
$\gamma$-Sn vanished \tb{when the pressure was decreased to} 3.5 GPa.

\begin{figure}[htb!]
\centering\includegraphics[clip,width=1\columnwidth]{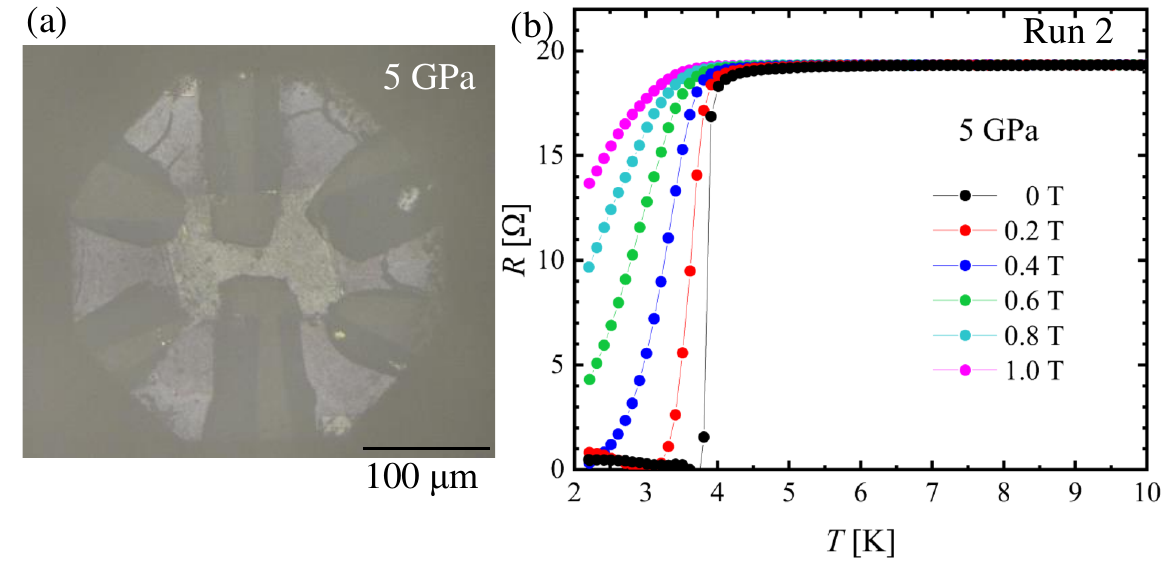}
\caption{\label{f_RT_P2} (Color online) (a) Optical image of the thin film of Sn \tb{at} 5 GPa and (b) $T$ dependence of $R$ at 5 GPa.}
\end{figure}

Figures~\ref{f_RT_H}(a) and (b) show \tb{the} $T$ dependence of $R$ \tb{for} the {thin film of Sn} under various $H$ at 2.5 and 10.5 GPa {\tb{under} non-hydrostatic pressure condition}.
$T_{\rm c}$ was observed \tb{at} 1 T, whereas the upper critical magnetic field ($H_{\rm c2}$) was approximately 0.1 T \tb{under} ambient pressure.
\tb{Drastic $H_{\rm c2}$ enhancement} was also observed \tb{in the} $\gamma$ phase at 10.5 GPa.
As shown in Fig.~\ref{f_RT_H}(c), $H_{\rm c2}$ estimated \tb{using} {Werthamer--Helfand--Hohenberg (WHH)} model reached several teslas \tb{at} high pressures \cite{helfandTemperaturePurityDependence1966, werthamerTemperaturePurityDependence1966, baumgartnerEffectsNeutronIrradiation2013}.
\tb{Notably, the $H_{\rm c2}$ enhancement of} $\beta$-Sn under \tb{the} {non-hydrostatic} pressure \tb{condition} was observed reproducibly.
Figure~\ref{f_RT_P2}(b) shows \tb{the} $T$ dependence of $R$ \tb{for} the {thin film of Sn} with the other setup \tb{at} 5 GPa.
The optical image of the {thin film of Sn} is shown in Fig.~\ref{f_RT_P2}(a).
The $R$ decrease suggesting the superconducting transition was observed even \tb{at} $H=1.0$ T.

\trr{There are two possible scenarios for the $H_{\rm c2}$ enhancement.
The first is the enhancement of the flux-pinning force.
The crystallographic defects introduced by non-hydrostatic pressure could serve as pinning centers, potentially leading to an increase in $H_{\rm c2}$.
The other is the shortening of the mean free path of the electrons and the resulting shortening of the coherence length.
Considering that the thickness of the thin film was approximately 100 nm, the electrical resistivity of the normal state near $T_{\rm c}$ was estimated to be approximately $2.5~\mu \Omega \cdot \rm{cm}$ for ambient pressure measurement.
This value is consistent with those reported in previous studies \cite{bangCharacterizationSuperconductingSn2019, ohkumaNonreciprocalSupercurrentThin2023}.
In contrast, the electrical resistivities of the thin films for 2.5 GPa of Run 1 and 5 GPa of Run 2 were approximately $1.7 \times 10^3~\mu \Omega \cdot \rm{cm}$ and $81~\mu \Omega \cdot \rm{cm}$, respectively.
The increase in the electrical resistivity near $T_{\rm c}$ suggests that the mean free path of the electrons is shortened by application of pressure.}

\subsection{Quasi-hydrostatic pressure on thin film}
\begin{figure}[htb!]
\centering\includegraphics[clip,width=0.7\columnwidth]{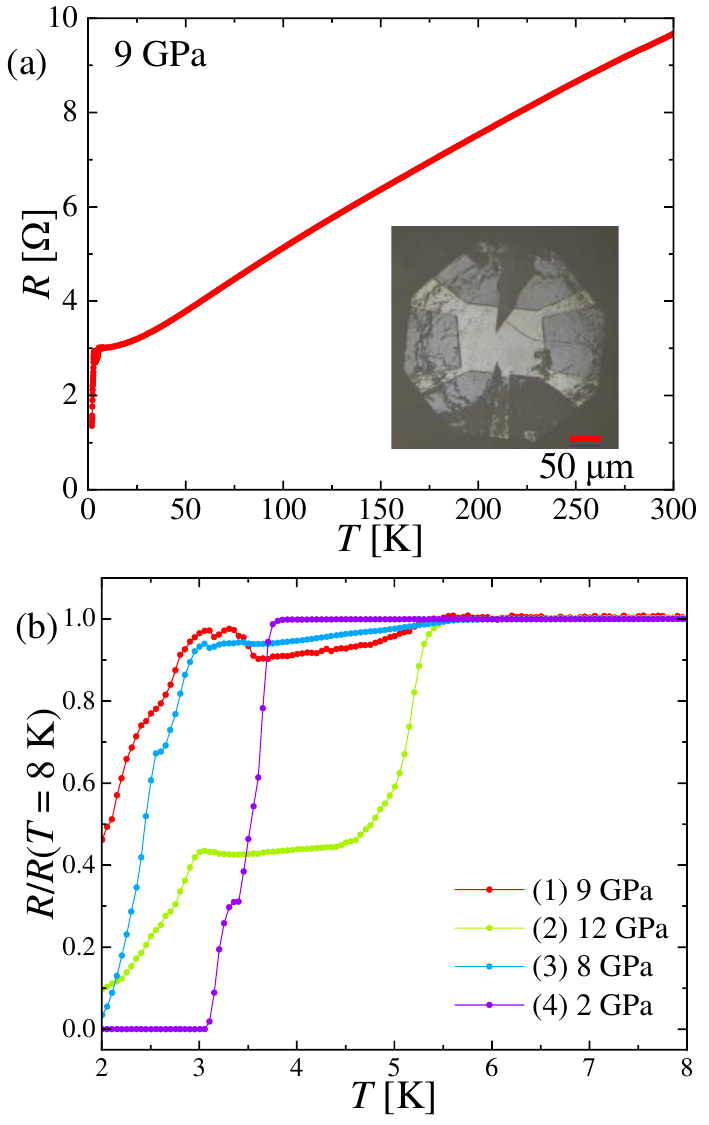}
\caption{\label{f_RT_Q} (Color online) (a) $T$ dependence of $R$ at 9 GPa.
The inset shows the optical \tb{image} of the thin film.
(b) $T$ dependence of $R$ under high pressures between 2 and 8 K.}
\end{figure}

{We performed \tb{high-pressure} generation with better hydrostatic pressure condition using \tb{a} liquid pressure transmitted medium, glycerol.
Figure~\ref{f_RT_Q}(a) shows \tb{the} $R$--$T$ \tb{of} the thin film at 9 GPa under quasi-hydrostatic pressure conditions.
The optical \tb{image} of the thin film inside \tb{the} DAC is shown in the inset of Fig.~\ref{f_RT_Q}(a).
The RRR was \tb{with in} 2--3 in quasi-hydrostatic pressure measurements, which was slightly higher than \tb{those for the} non-hydrostatic pressure condition.
The $R$ drop suggesting superconducting transition was observed around 5 K.
Figure~\ref{f_RT_Q}(b) shows $R$--$T$ under quasi-hydrostatic pressure \tb{condition} between 2 and 8 K with warming process. 
The $R$ \tb{value} was normalized \tb{using} the value at 8 K.
At 9 GPa, $R$ slightly decreased with decreasing $T$ around 5.3 K, suggesting the superconducting transition on $\gamma$-Sn.
The $R$ drop was also observed around 3 K.
The $\gamma$-Sn phase became significant \tb{on further pressure application}.
The onset of $T_{\rm c}$ slightly increased at 12 GPa.
\tb{We also} decreased the pressure \tb{from 12 to 8 GPa}.
The $R$ decrease suggesting the superconducting transition of $\gamma$-Sn was slightly observed, whereas the $\gamma$-Sn was clearly observed at 8.5 GPa \tb{under the non-hydrostatic pressure condition}.
\tb{With further pressure decrease}, $\gamma$-Sn vanished at 2 GPa.}

\section{Discussion}
\subsection{Pressure dependence of $T_{\rm c}$}
\begin{figure}[htb!]
\centering\includegraphics[clip,width=0.8\columnwidth]{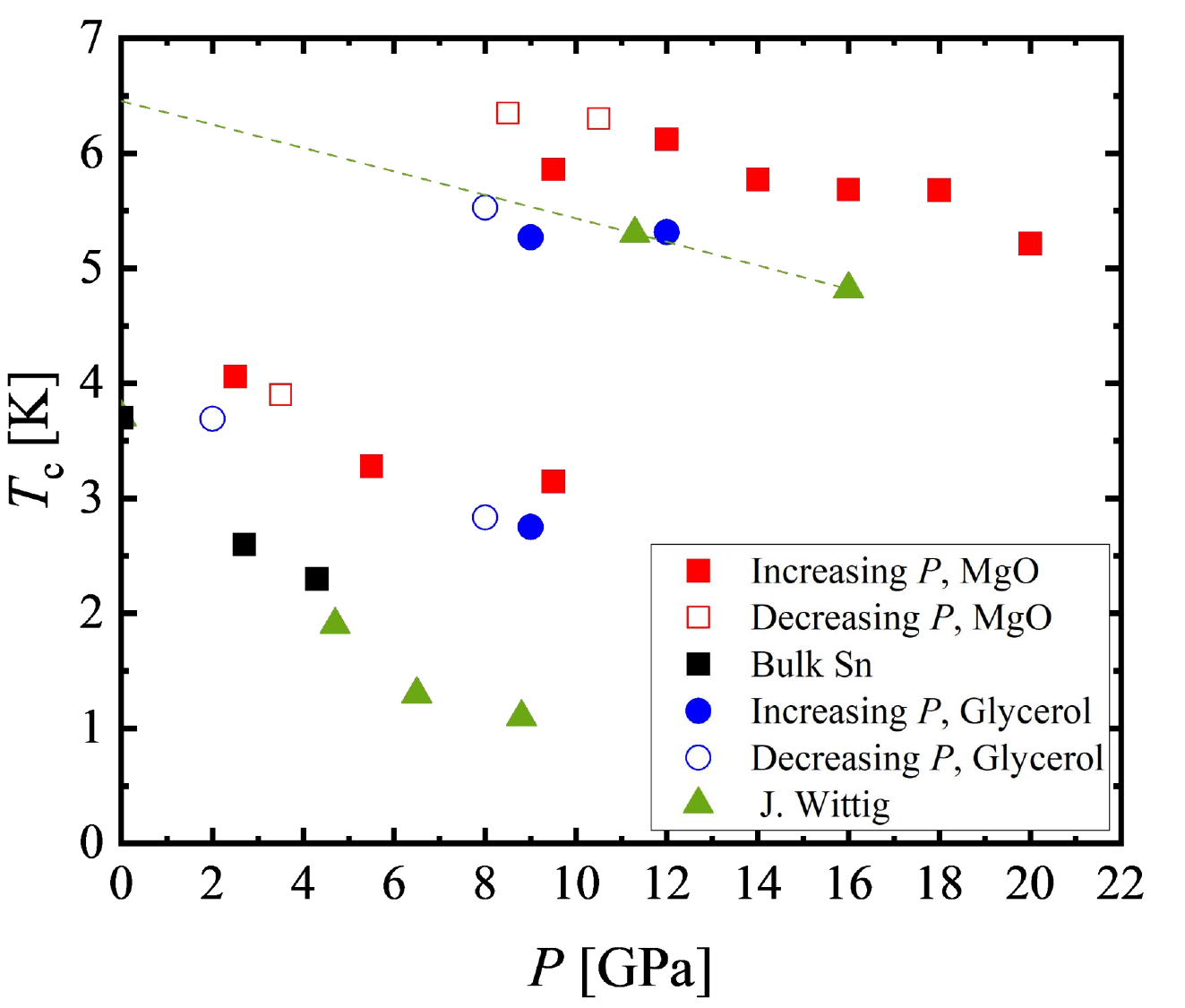}
\caption{\label{f_tc} (Color online) Pressure dependence of the superconducting transition temperature.
The green triangles indicate the results by Wittig \cite{wittigSupraleitungZinnUnd1966}.
\tr{The dotted line indicates a guide for the eyes.}}
\end{figure}

Figure~\ref{f_tc} shows the pressure dependence of $T_{\rm c}$ \tb{for} the thin film and bulk samples compared with \tb{results from a} previous study \cite{wittigSupraleitungZinnUnd1966}.
For the bulk Sn, the behavior of $T_{\rm c}$ \tb{with respect to} pressure was good agreement with the previous study.
\tb{A} similar tendency was observed \tb{for} the thin film in $\beta$-Sn phas;, however, \tb{its} $T_{\rm c}$ was higher \tb{compared to bulk Sn}.
One possible reason is the geometry of the thin film.
As shown in Fig~\ref{f1}(a), \tb{the thin film area} occupies approximately 70$\%$ of the culet of the diamond anvil, \tb{which produces} the pressure distribution.
In \tb{the} $\gamma$-Sn phase, $T_{\rm c}$ with quasi-hydrostatic pressure measurements showed \tb{a trend similar to} the previous results \cite{wittigSupraleitungZinnUnd1966}.
\tb{Meanwhile}, $T_{\rm c}$ with non-hydrostatic pressure showed \tb{a} higher value.
We observed the highest $T_{\rm c}$ of 6.3 K at 10.5 GPa, which was approximately 10$\%$ higher than that \tb{reported in} a previous study \cite{wittigSupraleitungZinnUnd1966}.
\tr{The highest $T_{\rm c}$ is not fully explained by the pressure gradient.
Assuming that the $T_{\rm c}$ of bulk $\gamma$-Sn continues to change at a rate of $-0.11$ K/GPa, it is necessary to decrease the pressure to 4 GPa from the $\gamma$ phase above 10 GPa.
However, the $\gamma$ phase cannot exist metastably under this pressure as was observed during the $\gamma$ to $\beta$ phase transition on the decompression process (Fig.~\ref{f_RT_P}(b)).}
{We {emphasize} that {we observed} the mid point of $T_{\rm c}$ \tb{to be} {greater than} 6.0 K (Fig.~\ref{f_RT_H}(b)).}

\subsection{Atomic force microscopy}
\begin{figure}[htb!]
\centering\includegraphics[clip,width=1\columnwidth]{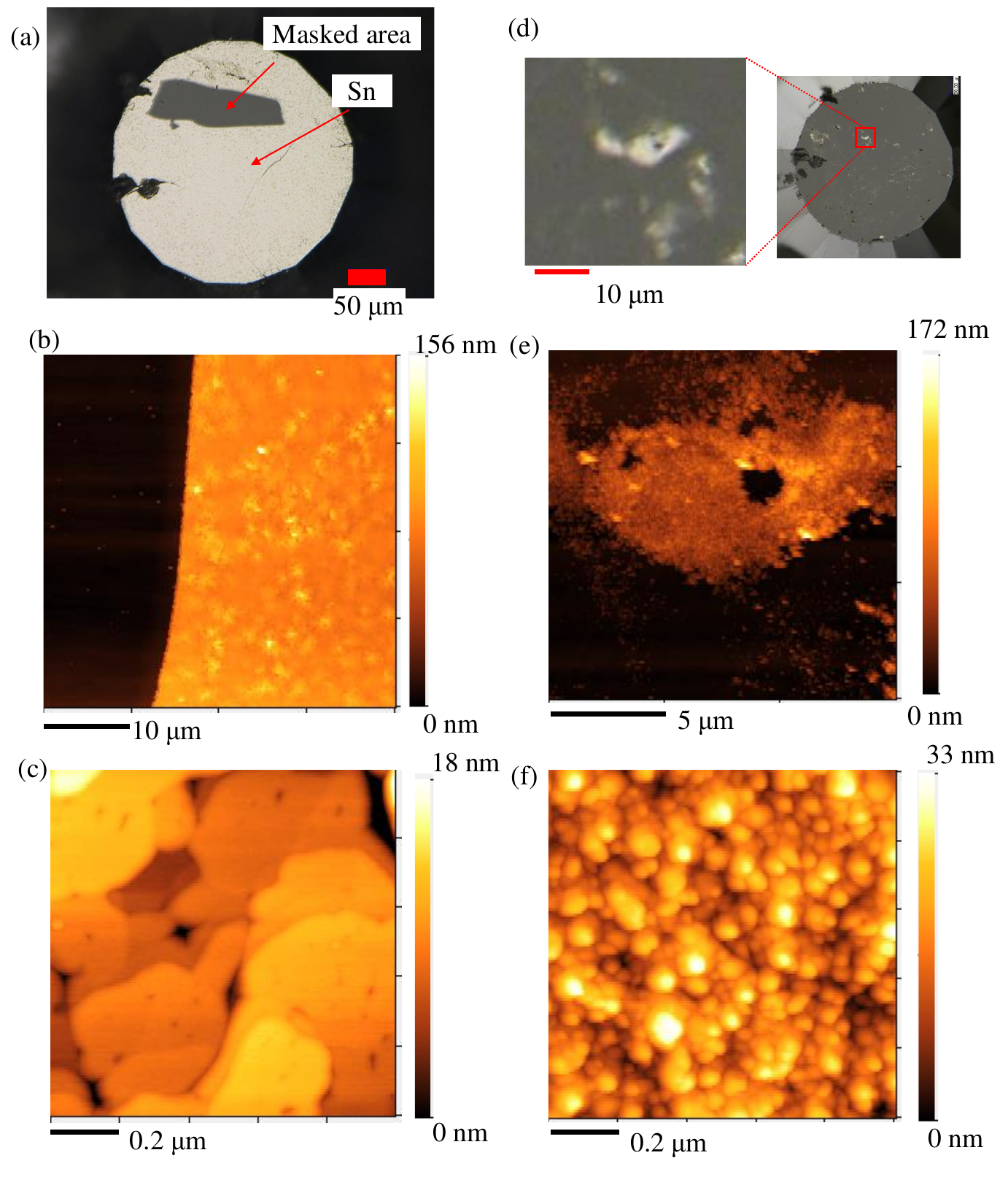}
\caption{\label{f_AFM1} (Color online) (a) Optical image of the thin film of Sn on a diamond anvil before \tb{pressurization}.
(b) and (c) AFM image on the thin film before \tb{pressurization}.
(d) Optical images of the thin film of Sn on the diamond anvil after \tb{pressurization}.
(e) and (f) AFM image of the thin film after \tb{pressurization}.
}
\end{figure}

To {investigate} the morphology on the thin films, we performed AFM measurements before and after applying pressure {\tb{under the} non-hydrostatic pressure condition}.
Figures~\ref{f_AFM1}(a) and (b) show the optical and AFM images of the thin film before \tb{pressurization}.
{The thin film of Sn was deposited on \tb{a} diamond anvil without BDD electrodes.}
The average \tb{film thickness} was evaluated to be 70 nm.
\tb{A magnified} AFM image is shown in Fig.~\ref{f_AFM1}(c).
The average grain size \tb{was} estimated to be approximately 300 nm.
Figure~\ref{f_AFM1}(d) shows the optical image of the thin film after applying pressure up to 4.2 GPa.
The pressure was evaluated by the diamond Raman shift.
Most of the thin film peeled off and was transferred to \tb{the MgO--epoxy} mixture.
We measured the small remaining \tb{thin-film area}.
Figure~\ref{f_AFM1}{(e)} shows the AFM image of the thin film after pressurization.
The average \tb{film thickness} was evaluated to be 60 nm.
\tb{A magnified} view is shown in Fig.~\ref{f_AFM1}{(f)}.
The grain size \tb{was} reduced to several tens of nanometers on pressure application.

\begin{figure}[htb!]
\centering\includegraphics[clip,width=1\columnwidth]{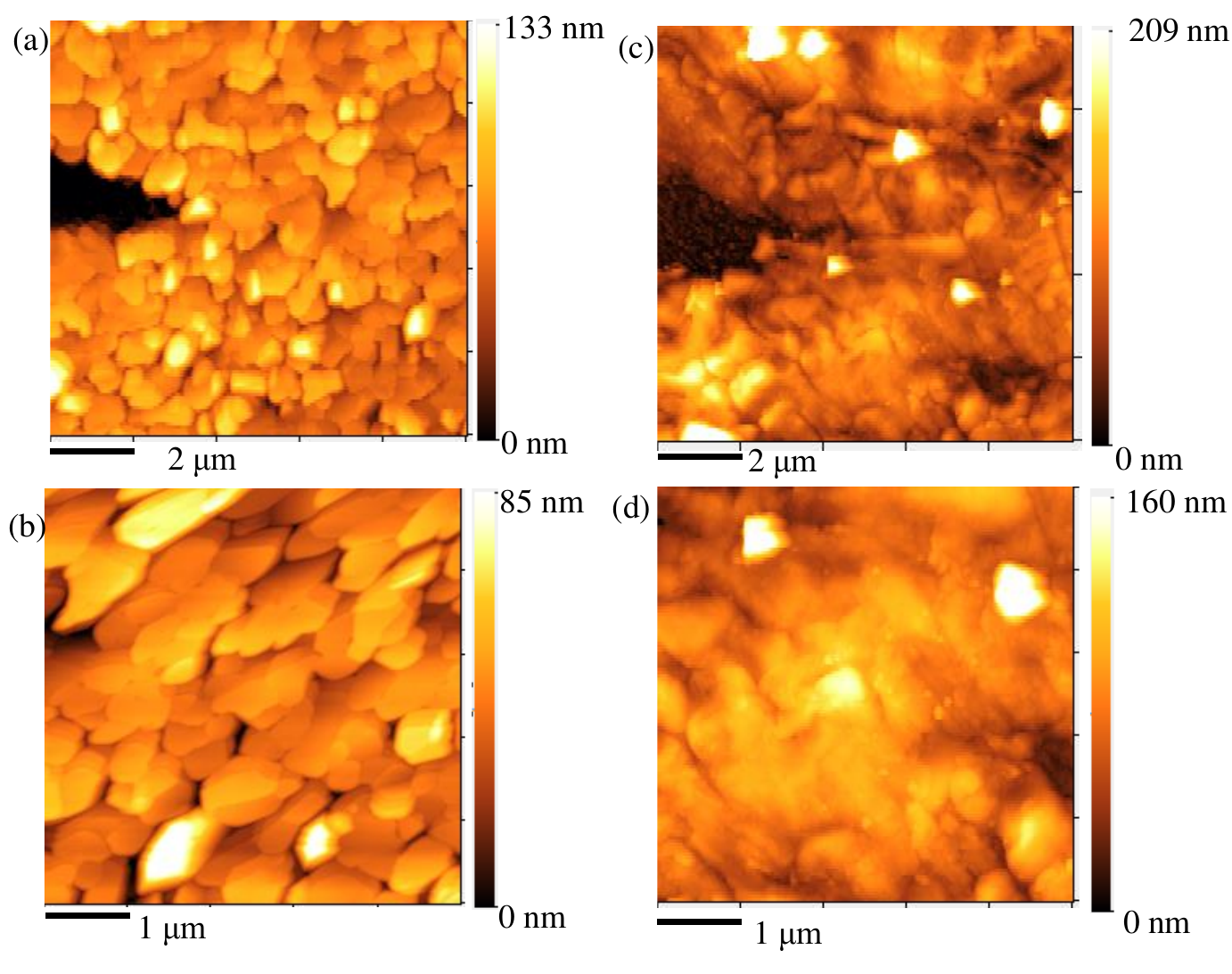}
\caption{\label{f_AFM2} (Color online) AFM \tb{under} quasi-hydrostatic pressure condition.
(a) and (b) AFM image of the thin film before \tb{pressurization}.
(c) and (d) AFM image of the thin film after \tb{pressurization}.
}
\end{figure}

{Next, we performed AFM measurements \tb{under} quasi-hydrostatic pressure \tb{condition for comparison with} the non-hydrostatic pressure measurements.
The thin film was \tb{the} same as that \tb{used in} the $R$--$T$ measurement.
Figures~\ref{f_AFM2}(a) and (b) show the AFM images \tb{of} the thin film before applying pressure.
The average thickness and average grain size were evaluated to be 82 and 600 nm, respectively.
Figures~\ref{f_AFM2}(c) and (d) show the AFM images on the thin film after applying pressure up to 12 GPa.
The \tb{pressure-transmitting} medium was removed \tb{using} ethanol and nitrogen gas.
The average thickness was evaluated to be 91 nm.
\tb{Unlike} the non-hydrostatic pressure measurements, grain refinement \tb{was} not observed.}

\tb{The} AFM results revealed \tb{that} the grain refinement was observed only \tb{under} the non-hydrostatic pressure condition.
In the $R$--$T$ results, the maximum $T_{\rm c}$ \tb{value under the} non-hydrostatic condition was 10$\%$ higher than that \tb{under the} quasi-hydrostatic condition, \tb{suggesting} that the grain refinement plays \tb{a pivotal} role \tb{in stabilizing} the higher $T_{\rm c}$.
\tr{Smaller grain sizes tend to have higher $T_{\rm c}$, because of phonon softening \tb{under} ambient pressure \cite{houbenInfluencePhononSoftening2020a}.
Houben $et~al.$ performed nuclear resonant inelastic x-ray scattering on nano-structured films and bulk Sn to investigate the phonon density of states and observed a decrease in \tb{the} high-energy phonon modes and a slight increase in \tb{the} low-energy phonon modes in nano-structured films \cite{houbenLatticeDynamicsSn2017a, houbenInfluencePhononSoftening2020a}.
Using the obtained phonon spectra, calculations based on the Allen--Dynes--McMillan formalism yielded $T_{\rm c}$ values in good agreement with the experimental data.
In nano-structured films, the electron--phonon coupling increased by up to 10$\%$, \tb{suggesting} that phonon softening and the associated change in electron--phonon coupling play a major role in \tb{the $T_{\rm c}$ increase}.
We consider that the $T_{\rm c}$ \tb{increase under the} non-hydrostatic pressure condition is related to the grain size \tb{and} that grain refinement could \tb{induce} to changes in electron-phonon coupling.}

\subsection{Possible higher $T_{\rm c}$ on Sn}
\begin{figure}[htb!]
\centering\includegraphics[clip,width=0.9\columnwidth]{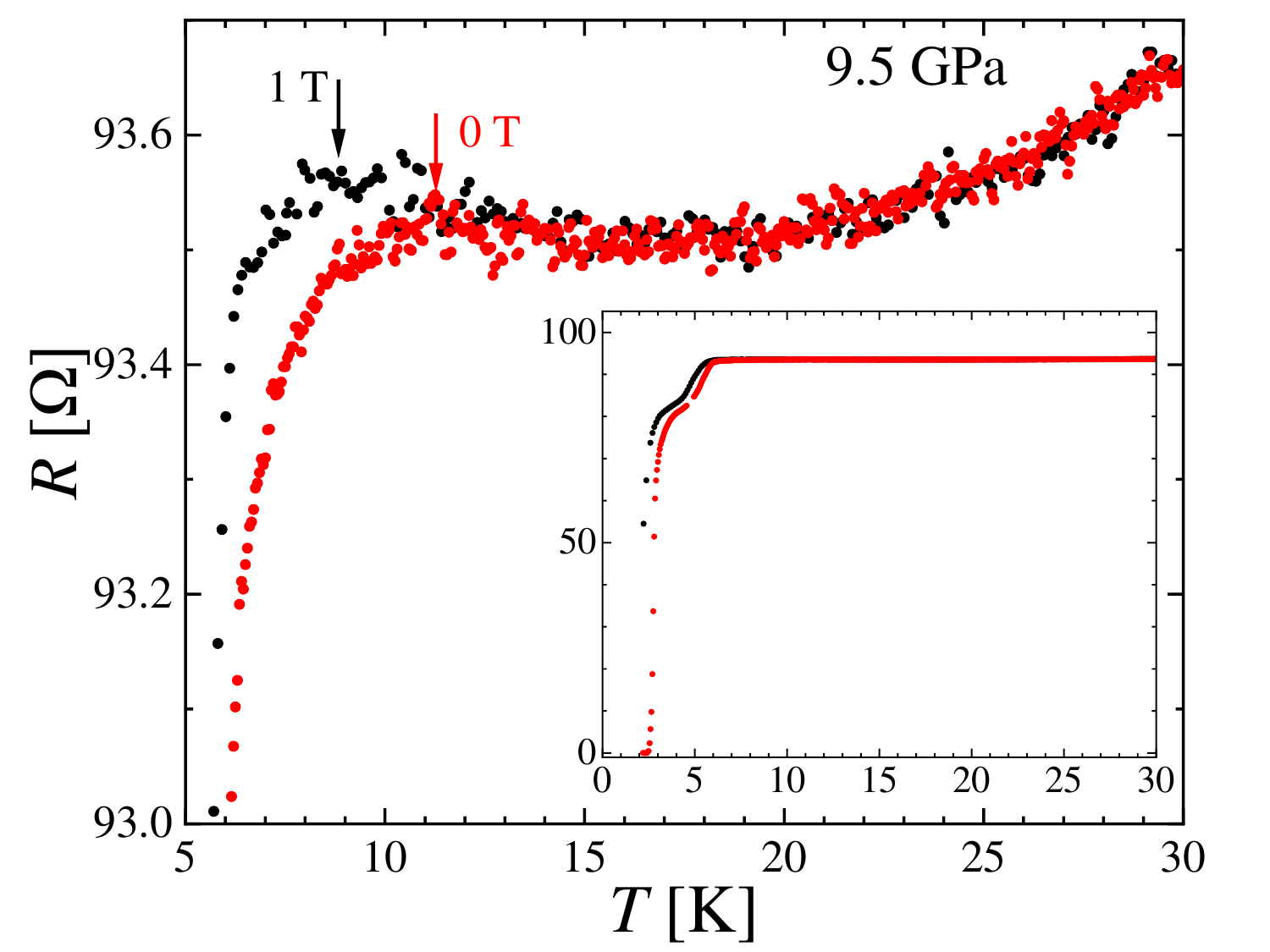}
\caption{\label{f4} (Color online) Temperature dependence of \tb{the} electrical resistance \tb{at} 9.5 GPa.
The downward arrows indicate the onset of the anomalies in the resistance.}
\end{figure}

Recently, anomalies \tb{in} magnetization, resistance, and heat capacity suggesting the superconducting transition \tb{were} observed around 5.5 K in nano-wires of Sn \cite{zhangDramaticEnhancementSuperconductivity2016}.
\tb{Further}, \tb{based on scanning tunneling microscopy}, the thin film of Sn \tb{deposited} on $\rm SrTiO_3$ substrate exhibited superconductivity around 8 K  \cite{shaoScanningTunnelingMicroscopic2018a}.
Indeed, in our thin film, the decrease of $R$ was observed at 11 K under 9.5 GPa, suggesting the signature of the superconducting transition, and the anomaly shifted to lower $T$ by applying the magnetic field (Fig~\ref{f4}).
On the other hand, some granular or amorphous thin films exhibit a resistance decrease at higher $T$ than $T_{\rm c}$ due to the effects of fluctuations \cite{gloverIdealResistiveTransition1967, aslamasovInfluenceFluctuationPairing1968, skocpolFluctuationsSuperconductingPhase1975, sonoraParaconductivityGranularFilms2019}.
Further investigations such as magnetic measurements \cite{hsiehImagingStressMagnetism2019, lesikMagneticMeasurementsMicrometersized2019}, heat capacity measurements \cite{umeoAlternatingCurrentCalorimeter2016}, and scanning tunneling microscopy \cite{caoSpectroscopicEvidenceSuperconductivity2023} under high pressure may \tb{offer} insights for the possible \tb{stabilization of} higher $T_{\rm c}$.

\section{Conclusion}
In conclusion, we demonstrated the pressure effect \tb{on} the superconductivity \tb{of} thin films of Sn.
We observed the superconductivity below {6.3 K} in \tb{the} $\gamma$-phase of Sn, which was approximately $10\%$ higher than {previous} bulk results.
\tb{Further}, the $H_{\rm c2}$ drastically increased under \tb{the} \tr{non-hydrostatic} high pressure \tb{condition}.
We also observed the signature of the superconducting transition at higher $T$ than $T_{\rm c}$.
AFM results suggest that the \tb{grain refinement under non-hydrostatic pressure contributes to the stabilization of the higher $T_{\rm c}$ of $\gamma$-Sn}.

\begin{acknowledgements}
{This work was supported by JSPS KAKENHI Grant No. 19H02177, 20H05644 and JST-Mirai Program Grant No. JPMJMI17A2.
This work was also supported by the Joint Usage/Research Center PRIUS, Ehime University, Japan}
\end{acknowledgements}

\section*{Data Availability Statement}
The data that support the findings of this article are openly available \cite{sasakiEnhancementSuperconductivityThin2025data} 
%
\bibliography{Sn1}

\begin{thebibliography}{67}%
\makeatletter
\providecommand \@ifxundefined [1]{%
 \@ifx{#1\undefined}
}%
\providecommand \@ifnum [1]{%
 \ifnum #1\expandafter \@firstoftwo
 \else \expandafter \@secondoftwo
 \fi
}%
\providecommand \@ifx [1]{%
 \ifx #1\expandafter \@firstoftwo
 \else \expandafter \@secondoftwo
 \fi
}%
\providecommand \natexlab [1]{#1}%
\providecommand \enquote  [1]{``#1''}%
\providecommand \bibnamefont  [1]{#1}%
\providecommand \bibfnamefont [1]{#1}%
\providecommand \citenamefont [1]{#1}%
\providecommand \href@noop [0]{\@secondoftwo}%
\providecommand \href [0]{\begingroup \@sanitize@url \@href}%
\providecommand \@href[1]{\@@startlink{#1}\@@href}%
\providecommand \@@href[1]{\endgroup#1\@@endlink}%
\providecommand \@sanitize@url [0]{\catcode `\\12\catcode `\$12\catcode
  `\&12\catcode `\#12\catcode `\^12\catcode `\_12\catcode `\%12\relax}%
\providecommand \@@startlink[1]{}%
\providecommand \@@endlink[0]{}%
\providecommand \url  [0]{\begingroup\@sanitize@url \@url }%
\providecommand \@url [1]{\endgroup\@href {#1}{\urlprefix }}%
\providecommand \urlprefix  [0]{URL }%
\providecommand \Eprint [0]{\href }%
\providecommand \doibase [0]{http://dx.doi.org/}%
\providecommand \selectlanguage [0]{\@gobble}%
\providecommand \bibinfo  [0]{\@secondoftwo}%
\providecommand \bibfield  [0]{\@secondoftwo}%
\providecommand \translation [1]{[#1]}%
\providecommand \BibitemOpen [0]{}%
\providecommand \bibitemStop [0]{}%
\providecommand \bibitemNoStop [0]{.\EOS\space}%
\providecommand \EOS [0]{\spacefactor3000\relax}%
\providecommand \BibitemShut  [1]{\csname bibitem#1\endcsname}%
\let\auto@bib@innerbib\@empty
\bibitem [{\citenamefont {Mao}\ \emph {et~al.}(2018)\citenamefont {Mao},
  \citenamefont {Chen}, \citenamefont {Ding}, \citenamefont {Li},\ and\
  \citenamefont {Wang}}]{maoSolidsLiquidsGases2018}%
  \BibitemOpen
  \bibfield  {author} {\bibinfo {author} {\bibfnamefont {H.-K.}\ \bibnamefont
  {Mao}}, \bibinfo {author} {\bibfnamefont {X.-J.}\ \bibnamefont {Chen}},
  \bibinfo {author} {\bibfnamefont {Y.}~\bibnamefont {Ding}}, \bibinfo {author}
  {\bibfnamefont {B.}~\bibnamefont {Li}}, \ and\ \bibinfo {author}
  {\bibfnamefont {L.}~\bibnamefont {Wang}},\ }\href {\doibase
  10.1103/RevModPhys.90.015007} {\bibfield  {journal} {\bibinfo  {journal}
  {Rev. Mod. Phys.}\ }\textbf {\bibinfo {volume} {90}},\ \bibinfo {pages}
  {015007} (\bibinfo {year} {2018})}\BibitemShut {NoStop}%
\bibitem [{\citenamefont
  {Yamanaka}(2010)}]{yamanakaSiliconClathratesCarbon2010}%
  \BibitemOpen
  \bibfield  {author} {\bibinfo {author} {\bibfnamefont {S.}~\bibnamefont
  {Yamanaka}},\ }\href {\doibase 10.1039/B918480E} {\bibfield  {journal}
  {\bibinfo  {journal} {Dalton Trans.}\ }\textbf {\bibinfo {volume} {39}},\
  \bibinfo {pages} {1901} (\bibinfo {year} {2010})}\BibitemShut {NoStop}%
\bibitem [{\citenamefont {Shimizu}\ \emph {et~al.}(1998)\citenamefont
  {Shimizu}, \citenamefont {Suhara}, \citenamefont {Ikumo}, \citenamefont
  {Eremets},\ and\ \citenamefont
  {Amaya}}]{shimizuSuperconductivityOxygen1998a}%
  \BibitemOpen
  \bibfield  {author} {\bibinfo {author} {\bibfnamefont {K.}~\bibnamefont
  {Shimizu}}, \bibinfo {author} {\bibfnamefont {K.}~\bibnamefont {Suhara}},
  \bibinfo {author} {\bibfnamefont {M.}~\bibnamefont {Ikumo}}, \bibinfo
  {author} {\bibfnamefont {M.~I.}\ \bibnamefont {Eremets}}, \ and\ \bibinfo
  {author} {\bibfnamefont {K.}~\bibnamefont {Amaya}},\ }\href {\doibase
  10.1038/31656} {\bibfield  {journal} {\bibinfo  {journal} {Nature}\ }\textbf
  {\bibinfo {volume} {393}},\ \bibinfo {pages} {767} (\bibinfo {year}
  {1998})}\BibitemShut {NoStop}%
\bibitem [{\citenamefont {Shimizu}\ \emph {et~al.}(2005)\citenamefont
  {Shimizu}, \citenamefont {Amaya},\ and\ \citenamefont
  {Suzuki}}]{shimizuPressureinducedSuperconductivityElemental2005}%
  \BibitemOpen
  \bibfield  {author} {\bibinfo {author} {\bibfnamefont {K.}~\bibnamefont
  {Shimizu}}, \bibinfo {author} {\bibfnamefont {K.}~\bibnamefont {Amaya}}, \
  and\ \bibinfo {author} {\bibfnamefont {N.}~\bibnamefont {Suzuki}},\ }\href
  {\doibase 10.1143/JPSJ.74.1345} {\bibfield  {journal} {\bibinfo  {journal}
  {J. Phys. Soc. Jpn.}\ }\textbf {\bibinfo {volume} {74}},\ \bibinfo {pages}
  {1345} (\bibinfo {year} {2005})}\BibitemShut {NoStop}%
\bibitem [{\citenamefont {Drozdov}\ \emph {et~al.}(2015)\citenamefont
  {Drozdov}, \citenamefont {Eremets}, \citenamefont {Troyan}, \citenamefont
  {Ksenofontov},\ and\ \citenamefont
  {Shylin}}]{drozdovConventionalSuperconductivity2032015}%
  \BibitemOpen
  \bibfield  {author} {\bibinfo {author} {\bibfnamefont {A.~P.}\ \bibnamefont
  {Drozdov}}, \bibinfo {author} {\bibfnamefont {M.~I.}\ \bibnamefont
  {Eremets}}, \bibinfo {author} {\bibfnamefont {I.~A.}\ \bibnamefont {Troyan}},
  \bibinfo {author} {\bibfnamefont {V.}~\bibnamefont {Ksenofontov}}, \ and\
  \bibinfo {author} {\bibfnamefont {S.~I.}\ \bibnamefont {Shylin}},\ }\href
  {\doibase 10.1038/nature14964} {\bibfield  {journal} {\bibinfo  {journal}
  {Nature}\ }\textbf {\bibinfo {volume} {525}},\ \bibinfo {pages} {73}
  (\bibinfo {year} {2015})}\BibitemShut {NoStop}%
\bibitem [{\citenamefont {Drozdov}\ \emph {et~al.}(2019)\citenamefont
  {Drozdov}, \citenamefont {Kong}, \citenamefont {Minkov}, \citenamefont
  {Besedin}, \citenamefont {Kuzovnikov}, \citenamefont {Mozaffari},
  \citenamefont {Balicas}, \citenamefont {Balakirev}, \citenamefont {Graf},
  \citenamefont {Prakapenka}, \citenamefont {Greenberg}, \citenamefont
  {Knyazev}, \citenamefont {Tkacz},\ and\ \citenamefont
  {Eremets}}]{drozdovSuperconductivity250Lanthanum2019}%
  \BibitemOpen
  \bibfield  {author} {\bibinfo {author} {\bibfnamefont {A.~P.}\ \bibnamefont
  {Drozdov}}, \bibinfo {author} {\bibfnamefont {P.~P.}\ \bibnamefont {Kong}},
  \bibinfo {author} {\bibfnamefont {V.~S.}\ \bibnamefont {Minkov}}, \bibinfo
  {author} {\bibfnamefont {S.~P.}\ \bibnamefont {Besedin}}, \bibinfo {author}
  {\bibfnamefont {M.~A.}\ \bibnamefont {Kuzovnikov}}, \bibinfo {author}
  {\bibfnamefont {S.}~\bibnamefont {Mozaffari}}, \bibinfo {author}
  {\bibfnamefont {L.}~\bibnamefont {Balicas}}, \bibinfo {author} {\bibfnamefont
  {F.~F.}\ \bibnamefont {Balakirev}}, \bibinfo {author} {\bibfnamefont {D.~E.}\
  \bibnamefont {Graf}}, \bibinfo {author} {\bibfnamefont {V.~B.}\ \bibnamefont
  {Prakapenka}}, \bibinfo {author} {\bibfnamefont {E.}~\bibnamefont
  {Greenberg}}, \bibinfo {author} {\bibfnamefont {D.~A.}\ \bibnamefont
  {Knyazev}}, \bibinfo {author} {\bibfnamefont {M.}~\bibnamefont {Tkacz}}, \
  and\ \bibinfo {author} {\bibfnamefont {M.~I.}\ \bibnamefont {Eremets}},\
  }\href {\doibase 10.1038/s41586-019-1201-8} {\bibfield  {journal} {\bibinfo
  {journal} {Nature}\ }\textbf {\bibinfo {volume} {569}},\ \bibinfo {pages}
  {528} (\bibinfo {year} {2019})}\BibitemShut {NoStop}%
\bibitem [{\citenamefont {Somayazulu}\ \emph {et~al.}(2019)\citenamefont
  {Somayazulu}, \citenamefont {Ahart}, \citenamefont {Mishra}, \citenamefont
  {Geballe}, \citenamefont {Baldini}, \citenamefont {Meng}, \citenamefont
  {Struzhkin},\ and\ \citenamefont
  {Hemley}}]{somayazuluEvidenceSuperconductivity2602019}%
  \BibitemOpen
  \bibfield  {author} {\bibinfo {author} {\bibfnamefont {M.}~\bibnamefont
  {Somayazulu}}, \bibinfo {author} {\bibfnamefont {M.}~\bibnamefont {Ahart}},
  \bibinfo {author} {\bibfnamefont {A.~K.}\ \bibnamefont {Mishra}}, \bibinfo
  {author} {\bibfnamefont {Z.~M.}\ \bibnamefont {Geballe}}, \bibinfo {author}
  {\bibfnamefont {M.}~\bibnamefont {Baldini}}, \bibinfo {author} {\bibfnamefont
  {Y.}~\bibnamefont {Meng}}, \bibinfo {author} {\bibfnamefont {V.~V.}\
  \bibnamefont {Struzhkin}}, \ and\ \bibinfo {author} {\bibfnamefont {R.~J.}\
  \bibnamefont {Hemley}},\ }\href {\doibase 10.1103/PhysRevLett.122.027001}
  {\bibfield  {journal} {\bibinfo  {journal} {Phys. Rev. Lett.}\ }\textbf
  {\bibinfo {volume} {122}},\ \bibinfo {pages} {027001} (\bibinfo {year}
  {2019})}\BibitemShut {NoStop}%
\bibitem [{\citenamefont {Matsumoto}\ \emph {et~al.}(2020)\citenamefont
  {Matsumoto}, \citenamefont {Einaga}, \citenamefont {Adachi}, \citenamefont
  {Yamamoto}, \citenamefont {Irifune}, \citenamefont {Terashima}, \citenamefont
  {Takeya}, \citenamefont {Nakamoto}, \citenamefont {Shimizu},\ and\
  \citenamefont {Takano}}]{matsumotoElectricalTransportMeasurements2020}%
  \BibitemOpen
  \bibfield  {author} {\bibinfo {author} {\bibfnamefont {R.}~\bibnamefont
  {Matsumoto}}, \bibinfo {author} {\bibfnamefont {M.}~\bibnamefont {Einaga}},
  \bibinfo {author} {\bibfnamefont {S.}~\bibnamefont {Adachi}}, \bibinfo
  {author} {\bibfnamefont {S.}~\bibnamefont {Yamamoto}}, \bibinfo {author}
  {\bibfnamefont {T.}~\bibnamefont {Irifune}}, \bibinfo {author} {\bibfnamefont
  {K.}~\bibnamefont {Terashima}}, \bibinfo {author} {\bibfnamefont
  {H.}~\bibnamefont {Takeya}}, \bibinfo {author} {\bibfnamefont
  {Y.}~\bibnamefont {Nakamoto}}, \bibinfo {author} {\bibfnamefont
  {K.}~\bibnamefont {Shimizu}}, \ and\ \bibinfo {author} {\bibfnamefont
  {Y.}~\bibnamefont {Takano}},\ }\href {\doibase 10.1088/1361-6668/abbdc5}
  {\bibfield  {journal} {\bibinfo  {journal} {Supercond. Sci. Technol.}\
  }\textbf {\bibinfo {volume} {33}},\ \bibinfo {pages} {124005} (\bibinfo
  {year} {2020})}\BibitemShut {NoStop}%
\bibitem [{\citenamefont {Semenok}\ \emph {et~al.}(2020)\citenamefont
  {Semenok}, \citenamefont {Kvashnin}, \citenamefont {Ivanova}, \citenamefont
  {Svitlyk}, \citenamefont {Fominski}, \citenamefont {Sadakov}, \citenamefont
  {Sobolevskiy}, \citenamefont {Pudalov}, \citenamefont {Troyan},\ and\
  \citenamefont {Oganov}}]{semenokSuperconductivity161Thorium2020}%
  \BibitemOpen
  \bibfield  {author} {\bibinfo {author} {\bibfnamefont {D.~V.}\ \bibnamefont
  {Semenok}}, \bibinfo {author} {\bibfnamefont {A.~G.}\ \bibnamefont
  {Kvashnin}}, \bibinfo {author} {\bibfnamefont {A.~G.}\ \bibnamefont
  {Ivanova}}, \bibinfo {author} {\bibfnamefont {V.}~\bibnamefont {Svitlyk}},
  \bibinfo {author} {\bibfnamefont {V.~Y.}\ \bibnamefont {Fominski}}, \bibinfo
  {author} {\bibfnamefont {A.~V.}\ \bibnamefont {Sadakov}}, \bibinfo {author}
  {\bibfnamefont {O.~A.}\ \bibnamefont {Sobolevskiy}}, \bibinfo {author}
  {\bibfnamefont {V.~M.}\ \bibnamefont {Pudalov}}, \bibinfo {author}
  {\bibfnamefont {I.~A.}\ \bibnamefont {Troyan}}, \ and\ \bibinfo {author}
  {\bibfnamefont {A.~R.}\ \bibnamefont {Oganov}},\ }\href {\doibase
  10.1016/j.mattod.2019.10.005} {\bibfield  {journal} {\bibinfo  {journal}
  {Materials Today}\ }\textbf {\bibinfo {volume} {33}},\ \bibinfo {pages} {36}
  (\bibinfo {year} {2020})}\BibitemShut {NoStop}%
\bibitem [{\citenamefont {Chen}\ \emph {et~al.}(2021)\citenamefont {Chen},
  \citenamefont {Semenok}, \citenamefont {Huang}, \citenamefont {Shu},
  \citenamefont {Li}, \citenamefont {Duan}, \citenamefont {Cui},\ and\
  \citenamefont {Oganov}}]{chenHighTemperatureSuperconductingPhases2021}%
  \BibitemOpen
  \bibfield  {author} {\bibinfo {author} {\bibfnamefont {W.}~\bibnamefont
  {Chen}}, \bibinfo {author} {\bibfnamefont {D.~V.}\ \bibnamefont {Semenok}},
  \bibinfo {author} {\bibfnamefont {X.}~\bibnamefont {Huang}}, \bibinfo
  {author} {\bibfnamefont {H.}~\bibnamefont {Shu}}, \bibinfo {author}
  {\bibfnamefont {X.}~\bibnamefont {Li}}, \bibinfo {author} {\bibfnamefont
  {D.}~\bibnamefont {Duan}}, \bibinfo {author} {\bibfnamefont {T.}~\bibnamefont
  {Cui}}, \ and\ \bibinfo {author} {\bibfnamefont {A.~R.}\ \bibnamefont
  {Oganov}},\ }\href {\doibase 10.1103/PhysRevLett.127.117001} {\bibfield
  {journal} {\bibinfo  {journal} {Phys. Rev. Lett.}\ }\textbf {\bibinfo
  {volume} {127}},\ \bibinfo {pages} {117001} (\bibinfo {year}
  {2021})}\BibitemShut {NoStop}%
\bibitem [{\citenamefont {Kong}\ \emph {et~al.}(2021)\citenamefont {Kong},
  \citenamefont {Minkov}, \citenamefont {Kuzovnikov}, \citenamefont {Drozdov},
  \citenamefont {Besedin}, \citenamefont {Mozaffari}, \citenamefont {Balicas},
  \citenamefont {Balakirev}, \citenamefont {Prakapenka}, \citenamefont
  {Chariton}, \citenamefont {Knyazev}, \citenamefont {Greenberg},\ and\
  \citenamefont {Eremets}}]{kongSuperconductivity243Yttriumhydrogen2021}%
  \BibitemOpen
  \bibfield  {author} {\bibinfo {author} {\bibfnamefont {P.}~\bibnamefont
  {Kong}}, \bibinfo {author} {\bibfnamefont {V.~S.}\ \bibnamefont {Minkov}},
  \bibinfo {author} {\bibfnamefont {M.~A.}\ \bibnamefont {Kuzovnikov}},
  \bibinfo {author} {\bibfnamefont {A.~P.}\ \bibnamefont {Drozdov}}, \bibinfo
  {author} {\bibfnamefont {S.~P.}\ \bibnamefont {Besedin}}, \bibinfo {author}
  {\bibfnamefont {S.}~\bibnamefont {Mozaffari}}, \bibinfo {author}
  {\bibfnamefont {L.}~\bibnamefont {Balicas}}, \bibinfo {author} {\bibfnamefont
  {F.~F.}\ \bibnamefont {Balakirev}}, \bibinfo {author} {\bibfnamefont {V.~B.}\
  \bibnamefont {Prakapenka}}, \bibinfo {author} {\bibfnamefont
  {S.}~\bibnamefont {Chariton}}, \bibinfo {author} {\bibfnamefont {D.~A.}\
  \bibnamefont {Knyazev}}, \bibinfo {author} {\bibfnamefont {E.}~\bibnamefont
  {Greenberg}}, \ and\ \bibinfo {author} {\bibfnamefont {M.~I.}\ \bibnamefont
  {Eremets}},\ }\href {\doibase 10.1038/s41467-021-25372-2} {\bibfield
  {journal} {\bibinfo  {journal} {Nat Commun}\ }\textbf {\bibinfo {volume}
  {12}},\ \bibinfo {pages} {5075} (\bibinfo {year} {2021})}\BibitemShut
  {NoStop}%
\bibitem [{\citenamefont {Troyan}\ \emph {et~al.}(2021)\citenamefont {Troyan},
  \citenamefont {Semenok}, \citenamefont {Kvashnin}, \citenamefont {Sadakov},
  \citenamefont {Sobolevskiy}, \citenamefont {Pudalov}, \citenamefont
  {Ivanova}, \citenamefont {Prakapenka}, \citenamefont {Greenberg},
  \citenamefont {Gavriliuk}, \citenamefont {Lyubutin}, \citenamefont
  {Struzhkin}, \citenamefont {Bergara}, \citenamefont {Errea}, \citenamefont
  {Bianco}, \citenamefont {Calandra}, \citenamefont {Mauri}, \citenamefont
  {Monacelli}, \citenamefont {Akashi},\ and\ \citenamefont
  {Oganov}}]{troyanAnomalousHighTemperatureSuperconductivity2021}%
  \BibitemOpen
  \bibfield  {author} {\bibinfo {author} {\bibfnamefont {I.~A.}\ \bibnamefont
  {Troyan}}, \bibinfo {author} {\bibfnamefont {D.~V.}\ \bibnamefont {Semenok}},
  \bibinfo {author} {\bibfnamefont {A.~G.}\ \bibnamefont {Kvashnin}}, \bibinfo
  {author} {\bibfnamefont {A.~V.}\ \bibnamefont {Sadakov}}, \bibinfo {author}
  {\bibfnamefont {O.~A.}\ \bibnamefont {Sobolevskiy}}, \bibinfo {author}
  {\bibfnamefont {V.~M.}\ \bibnamefont {Pudalov}}, \bibinfo {author}
  {\bibfnamefont {A.~G.}\ \bibnamefont {Ivanova}}, \bibinfo {author}
  {\bibfnamefont {V.~B.}\ \bibnamefont {Prakapenka}}, \bibinfo {author}
  {\bibfnamefont {E.}~\bibnamefont {Greenberg}}, \bibinfo {author}
  {\bibfnamefont {A.~G.}\ \bibnamefont {Gavriliuk}}, \bibinfo {author}
  {\bibfnamefont {I.~S.}\ \bibnamefont {Lyubutin}}, \bibinfo {author}
  {\bibfnamefont {V.~V.}\ \bibnamefont {Struzhkin}}, \bibinfo {author}
  {\bibfnamefont {A.}~\bibnamefont {Bergara}}, \bibinfo {author} {\bibfnamefont
  {I.}~\bibnamefont {Errea}}, \bibinfo {author} {\bibfnamefont
  {R.}~\bibnamefont {Bianco}}, \bibinfo {author} {\bibfnamefont
  {M.}~\bibnamefont {Calandra}}, \bibinfo {author} {\bibfnamefont
  {F.}~\bibnamefont {Mauri}}, \bibinfo {author} {\bibfnamefont
  {L.}~\bibnamefont {Monacelli}}, \bibinfo {author} {\bibfnamefont
  {R.}~\bibnamefont {Akashi}}, \ and\ \bibinfo {author} {\bibfnamefont {A.~R.}\
  \bibnamefont {Oganov}},\ }\href {\doibase 10.1002/adma.202006832} {\bibfield
  {journal} {\bibinfo  {journal} {Advanced Materials}\ }\textbf {\bibinfo
  {volume} {33}},\ \bibinfo {pages} {2006832} (\bibinfo {year}
  {2021})}\BibitemShut {NoStop}%
\bibitem [{\citenamefont {Li}\ \emph {et~al.}(2022)\citenamefont {Li},
  \citenamefont {He}, \citenamefont {Zhang}, \citenamefont {Wang},
  \citenamefont {Zhang}, \citenamefont {Jia}, \citenamefont {Feng},
  \citenamefont {Lu}, \citenamefont {Zhao}, \citenamefont {Zhang},
  \citenamefont {Min}, \citenamefont {Long}, \citenamefont {Yu}, \citenamefont
  {Wang}, \citenamefont {Ye}, \citenamefont {Zhang}, \citenamefont
  {Prakapenka}, \citenamefont {Chariton}, \citenamefont {Ginsberg},
  \citenamefont {Bass}, \citenamefont {Yuan}, \citenamefont {Liu},\ and\
  \citenamefont {Jin}}]{liSuperconductivity200Discovered2022}%
  \BibitemOpen
  \bibfield  {author} {\bibinfo {author} {\bibfnamefont {Z.}~\bibnamefont
  {Li}}, \bibinfo {author} {\bibfnamefont {X.}~\bibnamefont {He}}, \bibinfo
  {author} {\bibfnamefont {C.}~\bibnamefont {Zhang}}, \bibinfo {author}
  {\bibfnamefont {X.}~\bibnamefont {Wang}}, \bibinfo {author} {\bibfnamefont
  {S.}~\bibnamefont {Zhang}}, \bibinfo {author} {\bibfnamefont
  {Y.}~\bibnamefont {Jia}}, \bibinfo {author} {\bibfnamefont {S.}~\bibnamefont
  {Feng}}, \bibinfo {author} {\bibfnamefont {K.}~\bibnamefont {Lu}}, \bibinfo
  {author} {\bibfnamefont {J.}~\bibnamefont {Zhao}}, \bibinfo {author}
  {\bibfnamefont {J.}~\bibnamefont {Zhang}}, \bibinfo {author} {\bibfnamefont
  {B.}~\bibnamefont {Min}}, \bibinfo {author} {\bibfnamefont {Y.}~\bibnamefont
  {Long}}, \bibinfo {author} {\bibfnamefont {R.}~\bibnamefont {Yu}}, \bibinfo
  {author} {\bibfnamefont {L.}~\bibnamefont {Wang}}, \bibinfo {author}
  {\bibfnamefont {M.}~\bibnamefont {Ye}}, \bibinfo {author} {\bibfnamefont
  {Z.}~\bibnamefont {Zhang}}, \bibinfo {author} {\bibfnamefont
  {V.}~\bibnamefont {Prakapenka}}, \bibinfo {author} {\bibfnamefont
  {S.}~\bibnamefont {Chariton}}, \bibinfo {author} {\bibfnamefont {P.~A.}\
  \bibnamefont {Ginsberg}}, \bibinfo {author} {\bibfnamefont {J.}~\bibnamefont
  {Bass}}, \bibinfo {author} {\bibfnamefont {S.}~\bibnamefont {Yuan}}, \bibinfo
  {author} {\bibfnamefont {H.}~\bibnamefont {Liu}}, \ and\ \bibinfo {author}
  {\bibfnamefont {C.}~\bibnamefont {Jin}},\ }\href {\doibase
  10.1038/s41467-022-30454-w} {\bibfield  {journal} {\bibinfo  {journal} {Nat
  Commun}\ }\textbf {\bibinfo {volume} {13}},\ \bibinfo {pages} {2863}
  (\bibinfo {year} {2022})}\BibitemShut {NoStop}%
\bibitem [{\citenamefont {Ma}\ \emph {et~al.}(2022)\citenamefont {Ma},
  \citenamefont {Wang}, \citenamefont {Xie}, \citenamefont {Yang},
  \citenamefont {Wang}, \citenamefont {Zhou}, \citenamefont {Liu},
  \citenamefont {Yu}, \citenamefont {Zhao}, \citenamefont {Wang}, \citenamefont
  {Liu},\ and\ \citenamefont {Ma}}]{maHighTemperatureSuperconductingPhase2022}%
  \BibitemOpen
  \bibfield  {author} {\bibinfo {author} {\bibfnamefont {L.}~\bibnamefont
  {Ma}}, \bibinfo {author} {\bibfnamefont {K.}~\bibnamefont {Wang}}, \bibinfo
  {author} {\bibfnamefont {Y.}~\bibnamefont {Xie}}, \bibinfo {author}
  {\bibfnamefont {X.}~\bibnamefont {Yang}}, \bibinfo {author} {\bibfnamefont
  {Y.}~\bibnamefont {Wang}}, \bibinfo {author} {\bibfnamefont {M.}~\bibnamefont
  {Zhou}}, \bibinfo {author} {\bibfnamefont {H.}~\bibnamefont {Liu}}, \bibinfo
  {author} {\bibfnamefont {X.}~\bibnamefont {Yu}}, \bibinfo {author}
  {\bibfnamefont {Y.}~\bibnamefont {Zhao}}, \bibinfo {author} {\bibfnamefont
  {H.}~\bibnamefont {Wang}}, \bibinfo {author} {\bibfnamefont {G.}~\bibnamefont
  {Liu}}, \ and\ \bibinfo {author} {\bibfnamefont {Y.}~\bibnamefont {Ma}},\
  }\href {\doibase 10.1103/PhysRevLett.128.167001} {\bibfield  {journal}
  {\bibinfo  {journal} {Phys. Rev. Lett.}\ }\textbf {\bibinfo {volume} {128}},\
  \bibinfo {pages} {167001} (\bibinfo {year} {2022})}\BibitemShut {NoStop}%
\bibitem [{\citenamefont {Song}\ \emph {et~al.}(2023)\citenamefont {Song},
  \citenamefont {Bi}, \citenamefont {Nakamoto}, \citenamefont {Shimizu},
  \citenamefont {Liu}, \citenamefont {Zou}, \citenamefont {Liu}, \citenamefont
  {Wang},\ and\ \citenamefont
  {Ma}}]{songStoichiometricTernarySuperhydride2023}%
  \BibitemOpen
  \bibfield  {author} {\bibinfo {author} {\bibfnamefont {Y.}~\bibnamefont
  {Song}}, \bibinfo {author} {\bibfnamefont {J.}~\bibnamefont {Bi}}, \bibinfo
  {author} {\bibfnamefont {Y.}~\bibnamefont {Nakamoto}}, \bibinfo {author}
  {\bibfnamefont {K.}~\bibnamefont {Shimizu}}, \bibinfo {author} {\bibfnamefont
  {H.}~\bibnamefont {Liu}}, \bibinfo {author} {\bibfnamefont {B.}~\bibnamefont
  {Zou}}, \bibinfo {author} {\bibfnamefont {G.}~\bibnamefont {Liu}}, \bibinfo
  {author} {\bibfnamefont {H.}~\bibnamefont {Wang}}, \ and\ \bibinfo {author}
  {\bibfnamefont {Y.}~\bibnamefont {Ma}},\ }\href {\doibase
  10.1103/PhysRevLett.130.266001} {\bibfield  {journal} {\bibinfo  {journal}
  {Phys. Rev. Lett.}\ }\textbf {\bibinfo {volume} {130}},\ \bibinfo {pages}
  {266001} (\bibinfo {year} {2023})}\BibitemShut {NoStop}%
\bibitem [{\citenamefont {Cross}\ \emph {et~al.}(2024)\citenamefont {Cross},
  \citenamefont {Buhot}, \citenamefont {Brooks}, \citenamefont {Thomas},
  \citenamefont {Kleppe}, \citenamefont {Lord},\ and\ \citenamefont
  {Friedemann}}]{crossHightemperatureSuperconductivity$mathrmLa_4mathrmH_23$2024}%
  \BibitemOpen
  \bibfield  {author} {\bibinfo {author} {\bibfnamefont {S.}~\bibnamefont
  {Cross}}, \bibinfo {author} {\bibfnamefont {J.}~\bibnamefont {Buhot}},
  \bibinfo {author} {\bibfnamefont {A.}~\bibnamefont {Brooks}}, \bibinfo
  {author} {\bibfnamefont {W.}~\bibnamefont {Thomas}}, \bibinfo {author}
  {\bibfnamefont {A.}~\bibnamefont {Kleppe}}, \bibinfo {author} {\bibfnamefont
  {O.}~\bibnamefont {Lord}}, \ and\ \bibinfo {author} {\bibfnamefont
  {S.}~\bibnamefont {Friedemann}},\ }\href {\doibase
  10.1103/PhysRevB.109.L020503} {\bibfield  {journal} {\bibinfo  {journal}
  {Phys. Rev. B}\ }\textbf {\bibinfo {volume} {109}},\ \bibinfo {pages}
  {L020503} (\bibinfo {year} {2024})}\BibitemShut {NoStop}%
\bibitem [{\citenamefont {Sun}\ \emph {et~al.}(2023)\citenamefont {Sun},
  \citenamefont {Huo}, \citenamefont {Hu}, \citenamefont {Li}, \citenamefont
  {Liu}, \citenamefont {Han}, \citenamefont {Tang}, \citenamefont {Mao},
  \citenamefont {Yang}, \citenamefont {Wang}, \citenamefont {Cheng},
  \citenamefont {Yao}, \citenamefont {Zhang},\ and\ \citenamefont
  {Wang}}]{sunSignaturesSuperconductivity802023}%
  \BibitemOpen
  \bibfield  {author} {\bibinfo {author} {\bibfnamefont {H.}~\bibnamefont
  {Sun}}, \bibinfo {author} {\bibfnamefont {M.}~\bibnamefont {Huo}}, \bibinfo
  {author} {\bibfnamefont {X.}~\bibnamefont {Hu}}, \bibinfo {author}
  {\bibfnamefont {J.}~\bibnamefont {Li}}, \bibinfo {author} {\bibfnamefont
  {Z.}~\bibnamefont {Liu}}, \bibinfo {author} {\bibfnamefont {Y.}~\bibnamefont
  {Han}}, \bibinfo {author} {\bibfnamefont {L.}~\bibnamefont {Tang}}, \bibinfo
  {author} {\bibfnamefont {Z.}~\bibnamefont {Mao}}, \bibinfo {author}
  {\bibfnamefont {P.}~\bibnamefont {Yang}}, \bibinfo {author} {\bibfnamefont
  {B.}~\bibnamefont {Wang}}, \bibinfo {author} {\bibfnamefont {J.}~\bibnamefont
  {Cheng}}, \bibinfo {author} {\bibfnamefont {D.-X.}\ \bibnamefont {Yao}},
  \bibinfo {author} {\bibfnamefont {G.-M.}\ \bibnamefont {Zhang}}, \ and\
  \bibinfo {author} {\bibfnamefont {M.}~\bibnamefont {Wang}},\ }\href {\doibase
  10.1038/s41586-023-06408-7} {\bibfield  {journal} {\bibinfo  {journal}
  {Nature}\ }\textbf {\bibinfo {volume} {621}},\ \bibinfo {pages} {493}
  (\bibinfo {year} {2023})}\BibitemShut {NoStop}%
\bibitem [{\citenamefont {Sakakibara}\ \emph {et~al.}(2024)\citenamefont
  {Sakakibara}, \citenamefont {Ochi}, \citenamefont {Nagata}, \citenamefont
  {Ueki}, \citenamefont {Sakurai}, \citenamefont {Matsumoto}, \citenamefont
  {Terashima}, \citenamefont {Hirose}, \citenamefont {Ohta}, \citenamefont
  {Kato}, \citenamefont {Takano},\ and\ \citenamefont
  {Kuroki}}]{sakakibaraTheoreticalAnalysisPossibility2024}%
  \BibitemOpen
  \bibfield  {author} {\bibinfo {author} {\bibfnamefont {H.}~\bibnamefont
  {Sakakibara}}, \bibinfo {author} {\bibfnamefont {M.}~\bibnamefont {Ochi}},
  \bibinfo {author} {\bibfnamefont {H.}~\bibnamefont {Nagata}}, \bibinfo
  {author} {\bibfnamefont {Y.}~\bibnamefont {Ueki}}, \bibinfo {author}
  {\bibfnamefont {H.}~\bibnamefont {Sakurai}}, \bibinfo {author} {\bibfnamefont
  {R.}~\bibnamefont {Matsumoto}}, \bibinfo {author} {\bibfnamefont
  {K.}~\bibnamefont {Terashima}}, \bibinfo {author} {\bibfnamefont
  {K.}~\bibnamefont {Hirose}}, \bibinfo {author} {\bibfnamefont
  {H.}~\bibnamefont {Ohta}}, \bibinfo {author} {\bibfnamefont {M.}~\bibnamefont
  {Kato}}, \bibinfo {author} {\bibfnamefont {Y.}~\bibnamefont {Takano}}, \ and\
  \bibinfo {author} {\bibfnamefont {K.}~\bibnamefont {Kuroki}},\ }\href
  {\doibase 10.1103/PhysRevB.109.144511} {\bibfield  {journal} {\bibinfo
  {journal} {Phys. Rev. B}\ }\textbf {\bibinfo {volume} {109}},\ \bibinfo
  {pages} {144511} (\bibinfo {year} {2024})}\BibitemShut {NoStop}%
\bibitem [{\citenamefont {Liu}\ \emph {et~al.}(2022)\citenamefont {Liu},
  \citenamefont {Jiang}, \citenamefont {Wang}, \citenamefont {Li},
  \citenamefont {Li}, \citenamefont {Zhang}, \citenamefont {Wang},
  \citenamefont {Li},\ and\ \citenamefont {Yang}}]{liu$T_c$236Robust2022}%
  \BibitemOpen
  \bibfield  {author} {\bibinfo {author} {\bibfnamefont {X.}~\bibnamefont
  {Liu}}, \bibinfo {author} {\bibfnamefont {P.}~\bibnamefont {Jiang}}, \bibinfo
  {author} {\bibfnamefont {Y.}~\bibnamefont {Wang}}, \bibinfo {author}
  {\bibfnamefont {M.}~\bibnamefont {Li}}, \bibinfo {author} {\bibfnamefont
  {N.}~\bibnamefont {Li}}, \bibinfo {author} {\bibfnamefont {Q.}~\bibnamefont
  {Zhang}}, \bibinfo {author} {\bibfnamefont {Y.}~\bibnamefont {Wang}},
  \bibinfo {author} {\bibfnamefont {Y.-L.}\ \bibnamefont {Li}}, \ and\ \bibinfo
  {author} {\bibfnamefont {W.}~\bibnamefont {Yang}},\ }\href {\doibase
  10.1103/PhysRevB.105.224511} {\bibfield  {journal} {\bibinfo  {journal}
  {Phys. Rev. B}\ }\textbf {\bibinfo {volume} {105}},\ \bibinfo {pages}
  {224511} (\bibinfo {year} {2022})}\BibitemShut {NoStop}%
\bibitem [{\citenamefont {Zhang}\ \emph {et~al.}(2022)\citenamefont {Zhang},
  \citenamefont {He}, \citenamefont {Liu}, \citenamefont {Li}, \citenamefont
  {Lu}, \citenamefont {Zhang}, \citenamefont {Feng}, \citenamefont {Wang},
  \citenamefont {Peng}, \citenamefont {Long}, \citenamefont {Yu}, \citenamefont
  {Wang}, \citenamefont {Prakapenka}, \citenamefont {Chariton}, \citenamefont
  {Li}, \citenamefont {Liu}, \citenamefont {Chen},\ and\ \citenamefont
  {Jin}}]{zhangRecordHighTc2022}%
  \BibitemOpen
  \bibfield  {author} {\bibinfo {author} {\bibfnamefont {C.}~\bibnamefont
  {Zhang}}, \bibinfo {author} {\bibfnamefont {X.}~\bibnamefont {He}}, \bibinfo
  {author} {\bibfnamefont {C.}~\bibnamefont {Liu}}, \bibinfo {author}
  {\bibfnamefont {Z.}~\bibnamefont {Li}}, \bibinfo {author} {\bibfnamefont
  {K.}~\bibnamefont {Lu}}, \bibinfo {author} {\bibfnamefont {S.}~\bibnamefont
  {Zhang}}, \bibinfo {author} {\bibfnamefont {S.}~\bibnamefont {Feng}},
  \bibinfo {author} {\bibfnamefont {X.}~\bibnamefont {Wang}}, \bibinfo {author}
  {\bibfnamefont {Y.}~\bibnamefont {Peng}}, \bibinfo {author} {\bibfnamefont
  {Y.}~\bibnamefont {Long}}, \bibinfo {author} {\bibfnamefont {R.}~\bibnamefont
  {Yu}}, \bibinfo {author} {\bibfnamefont {L.}~\bibnamefont {Wang}}, \bibinfo
  {author} {\bibfnamefont {V.}~\bibnamefont {Prakapenka}}, \bibinfo {author}
  {\bibfnamefont {S.}~\bibnamefont {Chariton}}, \bibinfo {author}
  {\bibfnamefont {Q.}~\bibnamefont {Li}}, \bibinfo {author} {\bibfnamefont
  {H.}~\bibnamefont {Liu}}, \bibinfo {author} {\bibfnamefont {C.}~\bibnamefont
  {Chen}}, \ and\ \bibinfo {author} {\bibfnamefont {C.}~\bibnamefont {Jin}},\
  }\href {\doibase 10.1038/s41467-022-33077-3} {\bibfield  {journal} {\bibinfo
  {journal} {Nat Commun}\ }\textbf {\bibinfo {volume} {13}},\ \bibinfo {pages}
  {5411} (\bibinfo {year} {2022})}\BibitemShut {NoStop}%
\bibitem [{\citenamefont {Ying}\ \emph {et~al.}(2023)\citenamefont {Ying},
  \citenamefont {Liu}, \citenamefont {Lu}, \citenamefont {Wen}, \citenamefont
  {Gui}, \citenamefont {Zhang}, \citenamefont {Wang}, \citenamefont {Sun},\
  and\ \citenamefont {Chen}}]{yingRecordHigh362023}%
  \BibitemOpen
  \bibfield  {author} {\bibinfo {author} {\bibfnamefont {J.}~\bibnamefont
  {Ying}}, \bibinfo {author} {\bibfnamefont {S.}~\bibnamefont {Liu}}, \bibinfo
  {author} {\bibfnamefont {Q.}~\bibnamefont {Lu}}, \bibinfo {author}
  {\bibfnamefont {X.}~\bibnamefont {Wen}}, \bibinfo {author} {\bibfnamefont
  {Z.}~\bibnamefont {Gui}}, \bibinfo {author} {\bibfnamefont {Y.}~\bibnamefont
  {Zhang}}, \bibinfo {author} {\bibfnamefont {X.}~\bibnamefont {Wang}},
  \bibinfo {author} {\bibfnamefont {J.}~\bibnamefont {Sun}}, \ and\ \bibinfo
  {author} {\bibfnamefont {X.}~\bibnamefont {Chen}},\ }\href {\doibase
  10.1103/PhysRevLett.130.256002} {\bibfield  {journal} {\bibinfo  {journal}
  {Phys. Rev. Lett.}\ }\textbf {\bibinfo {volume} {130}},\ \bibinfo {pages}
  {256002} (\bibinfo {year} {2023})}\BibitemShut {NoStop}%
\bibitem [{\citenamefont {Buckel}\ and\ \citenamefont
  {Hilsch}(1954)}]{buckelEinflussKondensationBei1954a}%
  \BibitemOpen
  \bibfield  {author} {\bibinfo {author} {\bibfnamefont {W.}~\bibnamefont
  {Buckel}}\ and\ \bibinfo {author} {\bibfnamefont {R.}~\bibnamefont
  {Hilsch}},\ }\href {\doibase 10.1007/BF01337903} {\bibfield  {journal}
  {\bibinfo  {journal} {Z. Physik}\ }\textbf {\bibinfo {volume} {138}},\
  \bibinfo {pages} {109} (\bibinfo {year} {1954})}\BibitemShut {NoStop}%
\bibitem [{\citenamefont {Abeles}\ \emph {et~al.}(1966)\citenamefont {Abeles},
  \citenamefont {Cohen},\ and\ \citenamefont
  {Cullen}}]{abelesEnhancementSuperconductivityMetal1966}%
  \BibitemOpen
  \bibfield  {author} {\bibinfo {author} {\bibfnamefont {B.}~\bibnamefont
  {Abeles}}, \bibinfo {author} {\bibfnamefont {R.~W.}\ \bibnamefont {Cohen}}, \
  and\ \bibinfo {author} {\bibfnamefont {G.~W.}\ \bibnamefont {Cullen}},\
  }\href {\doibase 10.1103/PhysRevLett.17.632} {\bibfield  {journal} {\bibinfo
  {journal} {Phys. Rev. Lett.}\ }\textbf {\bibinfo {volume} {17}},\ \bibinfo
  {pages} {632} (\bibinfo {year} {1966})}\BibitemShut {NoStop}%
\bibitem [{\citenamefont {Garland}\ \emph {et~al.}(1968)\citenamefont
  {Garland}, \citenamefont {Bennemann},\ and\ \citenamefont
  {Mueller}}]{garlandEffectLatticeDisorder1968}%
  \BibitemOpen
  \bibfield  {author} {\bibinfo {author} {\bibfnamefont {J.~W.}\ \bibnamefont
  {Garland}}, \bibinfo {author} {\bibfnamefont {K.~H.}\ \bibnamefont
  {Bennemann}}, \ and\ \bibinfo {author} {\bibfnamefont {F.~M.}\ \bibnamefont
  {Mueller}},\ }\href {\doibase 10.1103/PhysRevLett.21.1315} {\bibfield
  {journal} {\bibinfo  {journal} {Phys. Rev. Lett.}\ }\textbf {\bibinfo
  {volume} {21}},\ \bibinfo {pages} {1315} (\bibinfo {year}
  {1968})}\BibitemShut {NoStop}%
\bibitem [{\citenamefont {Strongin}\ \emph {et~al.}(1968)\citenamefont
  {Strongin}, \citenamefont {Kammerer}, \citenamefont {Crow}, \citenamefont
  {Parks}, \citenamefont {Douglass},\ and\ \citenamefont
  {Jensen}}]{stronginEnhancedSuperconductivityLayered1968}%
  \BibitemOpen
  \bibfield  {author} {\bibinfo {author} {\bibfnamefont {M.}~\bibnamefont
  {Strongin}}, \bibinfo {author} {\bibfnamefont {O.~F.}\ \bibnamefont
  {Kammerer}}, \bibinfo {author} {\bibfnamefont {J.~E.}\ \bibnamefont {Crow}},
  \bibinfo {author} {\bibfnamefont {R.~D.}\ \bibnamefont {Parks}}, \bibinfo
  {author} {\bibfnamefont {D.~H.}\ \bibnamefont {Douglass}}, \ and\ \bibinfo
  {author} {\bibfnamefont {M.~A.}\ \bibnamefont {Jensen}},\ }\href {\doibase
  10.1103/PhysRevLett.21.1320} {\bibfield  {journal} {\bibinfo  {journal}
  {Phys. Rev. Lett.}\ }\textbf {\bibinfo {volume} {21}},\ \bibinfo {pages}
  {1320} (\bibinfo {year} {1968})}\BibitemShut {NoStop}%
\bibitem [{\citenamefont {Tian}\ \emph {et~al.}(2005)\citenamefont {Tian},
  \citenamefont {Wang}, \citenamefont {Kurtz}, \citenamefont {Liu},
  \citenamefont {Chan}, \citenamefont {Mayer},\ and\ \citenamefont
  {Mallouk}}]{tianDissipationQuasionedimensionalSuperconducting2005}%
  \BibitemOpen
  \bibfield  {author} {\bibinfo {author} {\bibfnamefont {M.}~\bibnamefont
  {Tian}}, \bibinfo {author} {\bibfnamefont {J.}~\bibnamefont {Wang}}, \bibinfo
  {author} {\bibfnamefont {J.~S.}\ \bibnamefont {Kurtz}}, \bibinfo {author}
  {\bibfnamefont {Y.}~\bibnamefont {Liu}}, \bibinfo {author} {\bibfnamefont
  {M.~H.~W.}\ \bibnamefont {Chan}}, \bibinfo {author} {\bibfnamefont {T.~S.}\
  \bibnamefont {Mayer}}, \ and\ \bibinfo {author} {\bibfnamefont {T.~E.}\
  \bibnamefont {Mallouk}},\ }\href {\doibase 10.1103/PhysRevB.71.104521}
  {\bibfield  {journal} {\bibinfo  {journal} {Phys. Rev. B}\ }\textbf {\bibinfo
  {volume} {71}},\ \bibinfo {pages} {104521} (\bibinfo {year}
  {2005})}\BibitemShut {NoStop}%
\bibitem [{\citenamefont {Beutel}\ \emph {et~al.}(2016)\citenamefont {Beutel},
  \citenamefont {Ebensperger}, \citenamefont {Thiemann}, \citenamefont
  {Untereiner}, \citenamefont {Fritz}, \citenamefont {Javaheri}, \citenamefont
  {N{\"a}gele}, \citenamefont {R{\"o}sslhuber}, \citenamefont {Dressel},\ and\
  \citenamefont {Scheffler}}]{beutelMicrowaveStudySuperconducting2016}%
  \BibitemOpen
  \bibfield  {author} {\bibinfo {author} {\bibfnamefont {M.~H.}\ \bibnamefont
  {Beutel}}, \bibinfo {author} {\bibfnamefont {N.~G.}\ \bibnamefont
  {Ebensperger}}, \bibinfo {author} {\bibfnamefont {M.}~\bibnamefont
  {Thiemann}}, \bibinfo {author} {\bibfnamefont {G.}~\bibnamefont
  {Untereiner}}, \bibinfo {author} {\bibfnamefont {V.}~\bibnamefont {Fritz}},
  \bibinfo {author} {\bibfnamefont {M.}~\bibnamefont {Javaheri}}, \bibinfo
  {author} {\bibfnamefont {J.}~\bibnamefont {N{\"a}gele}}, \bibinfo {author}
  {\bibfnamefont {R.}~\bibnamefont {R{\"o}sslhuber}}, \bibinfo {author}
  {\bibfnamefont {M.}~\bibnamefont {Dressel}}, \ and\ \bibinfo {author}
  {\bibfnamefont {M.}~\bibnamefont {Scheffler}},\ }\href {\doibase
  10.1088/0953-2048/29/8/085011} {\bibfield  {journal} {\bibinfo  {journal}
  {Supercond. Sci. Technol.}\ }\textbf {\bibinfo {volume} {29}},\ \bibinfo
  {pages} {085011} (\bibinfo {year} {2016})}\BibitemShut {NoStop}%
\bibitem [{\citenamefont {Bang}\ \emph {et~al.}(2019)\citenamefont {Bang},
  \citenamefont {Morrison}, \citenamefont {Rathnayaka}, \citenamefont
  {Lyuksyutov}, \citenamefont {Naugle},\ and\ \citenamefont
  {Teizer}}]{bangCharacterizationSuperconductingSn2019}%
  \BibitemOpen
  \bibfield  {author} {\bibinfo {author} {\bibfnamefont {W.}~\bibnamefont
  {Bang}}, \bibinfo {author} {\bibfnamefont {T.~D.}\ \bibnamefont {Morrison}},
  \bibinfo {author} {\bibfnamefont {K.~D.~D.}\ \bibnamefont {Rathnayaka}},
  \bibinfo {author} {\bibfnamefont {I.~F.}\ \bibnamefont {Lyuksyutov}},
  \bibinfo {author} {\bibfnamefont {D.~G.}\ \bibnamefont {Naugle}}, \ and\
  \bibinfo {author} {\bibfnamefont {W.}~\bibnamefont {Teizer}},\ }\href
  {\doibase 10.1016/j.tsf.2019.02.033} {\bibfield  {journal} {\bibinfo
  {journal} {Thin Solid Films}\ }\textbf {\bibinfo {volume} {676}},\ \bibinfo
  {pages} {138} (\bibinfo {year} {2019})}\BibitemShut {NoStop}%
\bibitem [{\citenamefont {Lozano}\ \emph {et~al.}(2019)\citenamefont {Lozano},
  \citenamefont {Couet}, \citenamefont {Petermann}, \citenamefont {Hamoir},
  \citenamefont {Jochum}, \citenamefont {Picot}, \citenamefont {Men{\'e}ndez},
  \citenamefont {Houben}, \citenamefont {Joly}, \citenamefont {Antohe},
  \citenamefont {Hu}, \citenamefont {Leu}, \citenamefont {Alatas},
  \citenamefont {Said}, \citenamefont {Roelants}, \citenamefont {Partoens},
  \citenamefont {Milo{\v s}evi{\'c}}, \citenamefont {Peeters}, \citenamefont
  {Piraux}, \citenamefont {{Van de Vondel}}, \citenamefont {Vantomme},
  \citenamefont {Temst},\ and\ \citenamefont
  {Van~Bael}}]{lozanoExperimentalObservationElectronphonon2019a}%
  \BibitemOpen
  \bibfield  {author} {\bibinfo {author} {\bibfnamefont {D.~P.}\ \bibnamefont
  {Lozano}}, \bibinfo {author} {\bibfnamefont {S.}~\bibnamefont {Couet}},
  \bibinfo {author} {\bibfnamefont {C.}~\bibnamefont {Petermann}}, \bibinfo
  {author} {\bibfnamefont {G.}~\bibnamefont {Hamoir}}, \bibinfo {author}
  {\bibfnamefont {J.~K.}\ \bibnamefont {Jochum}}, \bibinfo {author}
  {\bibfnamefont {T.}~\bibnamefont {Picot}}, \bibinfo {author} {\bibfnamefont
  {E.}~\bibnamefont {Men{\'e}ndez}}, \bibinfo {author} {\bibfnamefont
  {K.}~\bibnamefont {Houben}}, \bibinfo {author} {\bibfnamefont
  {V.}~\bibnamefont {Joly}}, \bibinfo {author} {\bibfnamefont {V.~A.}\
  \bibnamefont {Antohe}}, \bibinfo {author} {\bibfnamefont {M.~Y.}\
  \bibnamefont {Hu}}, \bibinfo {author} {\bibfnamefont {B.~M.}\ \bibnamefont
  {Leu}}, \bibinfo {author} {\bibfnamefont {A.}~\bibnamefont {Alatas}},
  \bibinfo {author} {\bibfnamefont {A.~H.}\ \bibnamefont {Said}}, \bibinfo
  {author} {\bibfnamefont {S.}~\bibnamefont {Roelants}}, \bibinfo {author}
  {\bibfnamefont {B.}~\bibnamefont {Partoens}}, \bibinfo {author}
  {\bibfnamefont {M.~V.}\ \bibnamefont {Milo{\v s}evi{\'c}}}, \bibinfo {author}
  {\bibfnamefont {F.~M.}\ \bibnamefont {Peeters}}, \bibinfo {author}
  {\bibfnamefont {L.}~\bibnamefont {Piraux}}, \bibinfo {author} {\bibfnamefont
  {J.}~\bibnamefont {{Van de Vondel}}}, \bibinfo {author} {\bibfnamefont
  {A.}~\bibnamefont {Vantomme}}, \bibinfo {author} {\bibfnamefont
  {K.}~\bibnamefont {Temst}}, \ and\ \bibinfo {author} {\bibfnamefont {M.~J.}\
  \bibnamefont {Van~Bael}},\ }\href {\doibase 10.1103/PhysRevB.99.064512}
  {\bibfield  {journal} {\bibinfo  {journal} {Phys. Rev. B}\ }\textbf {\bibinfo
  {volume} {99}},\ \bibinfo {pages} {064512} (\bibinfo {year}
  {2019})}\BibitemShut {NoStop}%
\bibitem [{\citenamefont {Knorr}\ and\ \citenamefont
  {Barth}(1970)}]{knorrSuperconductivityPhononSpectra1970}%
  \BibitemOpen
  \bibfield  {author} {\bibinfo {author} {\bibfnamefont {K.}~\bibnamefont
  {Knorr}}\ and\ \bibinfo {author} {\bibfnamefont {N.}~\bibnamefont {Barth}},\
  }\href {\doibase 10.1016/0038-1098(70)90265-6} {\bibfield  {journal}
  {\bibinfo  {journal} {Solid State Communications}\ }\textbf {\bibinfo
  {volume} {8}},\ \bibinfo {pages} {1085} (\bibinfo {year} {1970})}\BibitemShut
  {NoStop}%
\bibitem [{\citenamefont {Houben}\ \emph {et~al.}(2017)\citenamefont {Houben},
  \citenamefont {Couet}, \citenamefont {Trekels}, \citenamefont {Men{\'e}ndez},
  \citenamefont {Peissker}, \citenamefont {Seo}, \citenamefont {Hu},
  \citenamefont {Zhao}, \citenamefont {Alp}, \citenamefont {Roelants},
  \citenamefont {Partoens}, \citenamefont {Milo{\v s}evi{\'c}}, \citenamefont
  {Peeters}, \citenamefont {Bessas}, \citenamefont {Brown}, \citenamefont
  {Vantomme}, \citenamefont {Temst},\ and\ \citenamefont
  {Van~Bael}}]{houbenLatticeDynamicsSn2017a}%
  \BibitemOpen
  \bibfield  {author} {\bibinfo {author} {\bibfnamefont {K.}~\bibnamefont
  {Houben}}, \bibinfo {author} {\bibfnamefont {S.}~\bibnamefont {Couet}},
  \bibinfo {author} {\bibfnamefont {M.}~\bibnamefont {Trekels}}, \bibinfo
  {author} {\bibfnamefont {E.}~\bibnamefont {Men{\'e}ndez}}, \bibinfo {author}
  {\bibfnamefont {T.}~\bibnamefont {Peissker}}, \bibinfo {author}
  {\bibfnamefont {J.~W.}\ \bibnamefont {Seo}}, \bibinfo {author} {\bibfnamefont
  {M.~Y.}\ \bibnamefont {Hu}}, \bibinfo {author} {\bibfnamefont {J.~Y.}\
  \bibnamefont {Zhao}}, \bibinfo {author} {\bibfnamefont {E.~E.}\ \bibnamefont
  {Alp}}, \bibinfo {author} {\bibfnamefont {S.}~\bibnamefont {Roelants}},
  \bibinfo {author} {\bibfnamefont {B.}~\bibnamefont {Partoens}}, \bibinfo
  {author} {\bibfnamefont {M.~V.}\ \bibnamefont {Milo{\v s}evi{\'c}}}, \bibinfo
  {author} {\bibfnamefont {F.~M.}\ \bibnamefont {Peeters}}, \bibinfo {author}
  {\bibfnamefont {D.}~\bibnamefont {Bessas}}, \bibinfo {author} {\bibfnamefont
  {S.~A.}\ \bibnamefont {Brown}}, \bibinfo {author} {\bibfnamefont
  {A.}~\bibnamefont {Vantomme}}, \bibinfo {author} {\bibfnamefont
  {K.}~\bibnamefont {Temst}}, \ and\ \bibinfo {author} {\bibfnamefont {M.~J.}\
  \bibnamefont {Van~Bael}},\ }\href {\doibase 10.1103/PhysRevB.95.155413}
  {\bibfield  {journal} {\bibinfo  {journal} {Phys. Rev. B}\ }\textbf {\bibinfo
  {volume} {95}},\ \bibinfo {pages} {155413} (\bibinfo {year}
  {2017})}\BibitemShut {NoStop}%
\bibitem [{\citenamefont {Houben}\ \emph {et~al.}(2020)\citenamefont {Houben},
  \citenamefont {Jochum}, \citenamefont {Couet}, \citenamefont {Men{\'e}ndez},
  \citenamefont {Picot}, \citenamefont {Hu}, \citenamefont {Zhao},
  \citenamefont {Alp}, \citenamefont {Vantomme}, \citenamefont {Temst},\ and\
  \citenamefont {Van~Bael}}]{houbenInfluencePhononSoftening2020a}%
  \BibitemOpen
  \bibfield  {author} {\bibinfo {author} {\bibfnamefont {K.}~\bibnamefont
  {Houben}}, \bibinfo {author} {\bibfnamefont {J.~K.}\ \bibnamefont {Jochum}},
  \bibinfo {author} {\bibfnamefont {S.}~\bibnamefont {Couet}}, \bibinfo
  {author} {\bibfnamefont {E.}~\bibnamefont {Men{\'e}ndez}}, \bibinfo {author}
  {\bibfnamefont {T.}~\bibnamefont {Picot}}, \bibinfo {author} {\bibfnamefont
  {M.~Y.}\ \bibnamefont {Hu}}, \bibinfo {author} {\bibfnamefont {J.~Y.}\
  \bibnamefont {Zhao}}, \bibinfo {author} {\bibfnamefont {E.~E.}\ \bibnamefont
  {Alp}}, \bibinfo {author} {\bibfnamefont {A.}~\bibnamefont {Vantomme}},
  \bibinfo {author} {\bibfnamefont {K.}~\bibnamefont {Temst}}, \ and\ \bibinfo
  {author} {\bibfnamefont {M.~J.}\ \bibnamefont {Van~Bael}},\ }\href {\doibase
  10.1038/s41598-020-62617-4} {\bibfield  {journal} {\bibinfo  {journal} {Sci
  Rep}\ }\textbf {\bibinfo {volume} {10}},\ \bibinfo {pages} {5729} (\bibinfo
  {year} {2020})}\BibitemShut {NoStop}%
\bibitem [{\citenamefont {Barnett}\ \emph {et~al.}(2004)\citenamefont
  {Barnett}, \citenamefont {Bean},\ and\ \citenamefont
  {Hall}}]{barnettXRayDiffractionStudies2004}%
  \BibitemOpen
  \bibfield  {author} {\bibinfo {author} {\bibfnamefont {J.~D.}\ \bibnamefont
  {Barnett}}, \bibinfo {author} {\bibfnamefont {V.~E.}\ \bibnamefont {Bean}}, \
  and\ \bibinfo {author} {\bibfnamefont {H.~T.}\ \bibnamefont {Hall}},\ }\href
  {\doibase 10.1063/1.1708275} {\bibfield  {journal} {\bibinfo  {journal}
  {Journal of Applied Physics}\ }\textbf {\bibinfo {volume} {37}},\ \bibinfo
  {pages} {875} (\bibinfo {year} {2004})}\BibitemShut {NoStop}%
\bibitem [{\citenamefont {Salamat}\ \emph {et~al.}(2013)\citenamefont
  {Salamat}, \citenamefont {Briggs}, \citenamefont {Bouvier}, \citenamefont
  {Petitgirard}, \citenamefont {Dewaele}, \citenamefont {Cutler}, \citenamefont
  {Cor{\`a}}, \citenamefont {Daisenberger}, \citenamefont {Garbarino},\ and\
  \citenamefont {McMillan}}]{salamatHighpressureStructuralTransformations2013}%
  \BibitemOpen
  \bibfield  {author} {\bibinfo {author} {\bibfnamefont {A.}~\bibnamefont
  {Salamat}}, \bibinfo {author} {\bibfnamefont {R.}~\bibnamefont {Briggs}},
  \bibinfo {author} {\bibfnamefont {P.}~\bibnamefont {Bouvier}}, \bibinfo
  {author} {\bibfnamefont {S.}~\bibnamefont {Petitgirard}}, \bibinfo {author}
  {\bibfnamefont {A.}~\bibnamefont {Dewaele}}, \bibinfo {author} {\bibfnamefont
  {M.~E.}\ \bibnamefont {Cutler}}, \bibinfo {author} {\bibfnamefont
  {F.}~\bibnamefont {Cor{\`a}}}, \bibinfo {author} {\bibfnamefont
  {D.}~\bibnamefont {Daisenberger}}, \bibinfo {author} {\bibfnamefont
  {G.}~\bibnamefont {Garbarino}}, \ and\ \bibinfo {author} {\bibfnamefont
  {P.~F.}\ \bibnamefont {McMillan}},\ }\href {\doibase
  10.1103/PhysRevB.88.104104} {\bibfield  {journal} {\bibinfo  {journal} {Phys.
  Rev. B}\ }\textbf {\bibinfo {volume} {88}},\ \bibinfo {pages} {104104}
  (\bibinfo {year} {2013})}\BibitemShut {NoStop}%
\bibitem [{\citenamefont {Deffrennes}\ \emph {et~al.}(2022)\citenamefont
  {Deffrennes}, \citenamefont {Faure}, \citenamefont {Bottin}, \citenamefont
  {Joubert},\ and\ \citenamefont {Oudot}}]{deffrennesTinSnHigh2022}%
  \BibitemOpen
  \bibfield  {author} {\bibinfo {author} {\bibfnamefont {G.}~\bibnamefont
  {Deffrennes}}, \bibinfo {author} {\bibfnamefont {P.}~\bibnamefont {Faure}},
  \bibinfo {author} {\bibfnamefont {F.}~\bibnamefont {Bottin}}, \bibinfo
  {author} {\bibfnamefont {J.-M.}\ \bibnamefont {Joubert}}, \ and\ \bibinfo
  {author} {\bibfnamefont {B.}~\bibnamefont {Oudot}},\ }\href {\doibase
  10.1016/j.jallcom.2022.165675} {\bibfield  {journal} {\bibinfo  {journal}
  {Journal of Alloys and Compounds}\ }\textbf {\bibinfo {volume} {919}},\
  \bibinfo {pages} {165675} (\bibinfo {year} {2022})}\BibitemShut {NoStop}%
\bibitem [{\citenamefont {Wittig}(1966)}]{wittigSupraleitungZinnUnd1966}%
  \BibitemOpen
  \bibfield  {author} {\bibinfo {author} {\bibfnamefont {J.}~\bibnamefont
  {Wittig}},\ }\href {\doibase 10.1007/BF01328890} {\bibfield  {journal}
  {\bibinfo  {journal} {Z. Physik}\ }\textbf {\bibinfo {volume} {195}},\
  \bibinfo {pages} {228} (\bibinfo {year} {1966})}\BibitemShut {NoStop}%
\bibitem [{\citenamefont {Matsumoto}\ \emph {et~al.}(2016)\citenamefont
  {Matsumoto}, \citenamefont {Sasama}, \citenamefont {Fujioka}, \citenamefont
  {Irifune}, \citenamefont {Tanaka}, \citenamefont {Yamaguchi}, \citenamefont
  {Takeya},\ and\ \citenamefont {Takano}}]{matsumotoNoteNovelDiamond2016}%
  \BibitemOpen
  \bibfield  {author} {\bibinfo {author} {\bibfnamefont {R.}~\bibnamefont
  {Matsumoto}}, \bibinfo {author} {\bibfnamefont {Y.}~\bibnamefont {Sasama}},
  \bibinfo {author} {\bibfnamefont {M.}~\bibnamefont {Fujioka}}, \bibinfo
  {author} {\bibfnamefont {T.}~\bibnamefont {Irifune}}, \bibinfo {author}
  {\bibfnamefont {M.}~\bibnamefont {Tanaka}}, \bibinfo {author} {\bibfnamefont
  {T.}~\bibnamefont {Yamaguchi}}, \bibinfo {author} {\bibfnamefont
  {H.}~\bibnamefont {Takeya}}, \ and\ \bibinfo {author} {\bibfnamefont
  {Y.}~\bibnamefont {Takano}},\ }\href {\doibase 10.1063/1.4959154} {\bibfield
  {journal} {\bibinfo  {journal} {Review of Scientific Instruments}\ }\textbf
  {\bibinfo {volume} {87}},\ \bibinfo {pages} {076103} (\bibinfo {year}
  {2016})}\BibitemShut {NoStop}%
\bibitem [{\citenamefont {Matsumoto}\ \emph {et~al.}(2017)\citenamefont
  {Matsumoto}, \citenamefont {Irifune}, \citenamefont {Tanaka}, \citenamefont
  {Takeya},\ and\ \citenamefont {Takano}}]{matsumotoDiamondAnvilCell2017}%
  \BibitemOpen
  \bibfield  {author} {\bibinfo {author} {\bibfnamefont {R.}~\bibnamefont
  {Matsumoto}}, \bibinfo {author} {\bibfnamefont {T.}~\bibnamefont {Irifune}},
  \bibinfo {author} {\bibfnamefont {M.}~\bibnamefont {Tanaka}}, \bibinfo
  {author} {\bibfnamefont {H.}~\bibnamefont {Takeya}}, \ and\ \bibinfo {author}
  {\bibfnamefont {Y.}~\bibnamefont {Takano}},\ }\href {\doibase
  10.7567/JJAP.56.05FC01} {\bibfield  {journal} {\bibinfo  {journal} {Jpn. J.
  Appl. Phys.}\ }\textbf {\bibinfo {volume} {56}},\ \bibinfo {pages} {05FC01}
  (\bibinfo {year} {2017})}\BibitemShut {NoStop}%
\bibitem [{\citenamefont {Matsumoto}\ \emph {et~al.}(2018)\citenamefont
  {Matsumoto}, \citenamefont {Yamashita}, \citenamefont {Hara}, \citenamefont
  {Irifune}, \citenamefont {Adachi}, \citenamefont {Takeya},\ and\
  \citenamefont {Takano}}]{matsumotoDiamondAnvilCells2018}%
  \BibitemOpen
  \bibfield  {author} {\bibinfo {author} {\bibfnamefont {R.}~\bibnamefont
  {Matsumoto}}, \bibinfo {author} {\bibfnamefont {A.}~\bibnamefont
  {Yamashita}}, \bibinfo {author} {\bibfnamefont {H.}~\bibnamefont {Hara}},
  \bibinfo {author} {\bibfnamefont {T.}~\bibnamefont {Irifune}}, \bibinfo
  {author} {\bibfnamefont {S.}~\bibnamefont {Adachi}}, \bibinfo {author}
  {\bibfnamefont {H.}~\bibnamefont {Takeya}}, \ and\ \bibinfo {author}
  {\bibfnamefont {Y.}~\bibnamefont {Takano}},\ }\href {\doibase
  10.7567/APEX.11.053101} {\bibfield  {journal} {\bibinfo  {journal} {Appl.
  Phys. Express}\ }\textbf {\bibinfo {volume} {11}},\ \bibinfo {pages} {053101}
  (\bibinfo {year} {2018})}\BibitemShut {NoStop}%
\bibitem [{\citenamefont {Mao}\ \emph {et~al.}(1978)\citenamefont {Mao},
  \citenamefont {Bell}, \citenamefont {Shaner},\ and\ \citenamefont
  {Steinberg}}]{maoSpecificVolumeMeasurements2008}%
  \BibitemOpen
  \bibfield  {author} {\bibinfo {author} {\bibfnamefont {H.~K.}\ \bibnamefont
  {Mao}}, \bibinfo {author} {\bibfnamefont {P.~M.}\ \bibnamefont {Bell}},
  \bibinfo {author} {\bibfnamefont {J.~W.}\ \bibnamefont {Shaner}}, \ and\
  \bibinfo {author} {\bibfnamefont {D.~J.}\ \bibnamefont {Steinberg}},\ }\href
  {\doibase 10.1063/1.325277} {\bibfield  {journal} {\bibinfo  {journal}
  {Journal of Applied Physics}\ }\textbf {\bibinfo {volume} {49}},\ \bibinfo
  {pages} {3276} (\bibinfo {year} {1978})}\BibitemShut {NoStop}%
\bibitem [{\citenamefont {Akahama}\ and\ \citenamefont
  {Kawamura}(2004)}]{akahamaHighpressureRamanSpectroscopy2004}%
  \BibitemOpen
  \bibfield  {author} {\bibinfo {author} {\bibfnamefont {Y.}~\bibnamefont
  {Akahama}}\ and\ \bibinfo {author} {\bibfnamefont {H.}~\bibnamefont
  {Kawamura}},\ }\href {\doibase 10.1063/1.1778482} {\bibfield  {journal}
  {\bibinfo  {journal} {Journal of Applied Physics}\ }\textbf {\bibinfo
  {volume} {96}},\ \bibinfo {pages} {3748} (\bibinfo {year}
  {2004})}\BibitemShut {NoStop}%
\bibitem [{\citenamefont {Takano}\ \emph {et~al.}(2004)\citenamefont {Takano},
  \citenamefont {Nagao}, \citenamefont {Sakaguchi}, \citenamefont {Tachiki},
  \citenamefont {Hatano}, \citenamefont {Kobayashi}, \citenamefont {Umezawa},\
  and\ \citenamefont {Kawarada}}]{takanoSuperconductivityDiamondThin2004}%
  \BibitemOpen
  \bibfield  {author} {\bibinfo {author} {\bibfnamefont {Y.}~\bibnamefont
  {Takano}}, \bibinfo {author} {\bibfnamefont {M.}~\bibnamefont {Nagao}},
  \bibinfo {author} {\bibfnamefont {I.}~\bibnamefont {Sakaguchi}}, \bibinfo
  {author} {\bibfnamefont {M.}~\bibnamefont {Tachiki}}, \bibinfo {author}
  {\bibfnamefont {T.}~\bibnamefont {Hatano}}, \bibinfo {author} {\bibfnamefont
  {K.}~\bibnamefont {Kobayashi}}, \bibinfo {author} {\bibfnamefont
  {H.}~\bibnamefont {Umezawa}}, \ and\ \bibinfo {author} {\bibfnamefont
  {H.}~\bibnamefont {Kawarada}},\ }\href {\doibase 10.1063/1.1802389}
  {\bibfield  {journal} {\bibinfo  {journal} {Applied Physics Letters}\
  }\textbf {\bibinfo {volume} {85}},\ \bibinfo {pages} {2851} (\bibinfo {year}
  {2004})}\BibitemShut {NoStop}%
\bibitem [{\citenamefont {Adachi}\ \emph {et~al.}(2020)\citenamefont {Adachi},
  \citenamefont {Matsumoto}, \citenamefont {Yamamoto}, \citenamefont
  {Yamamoto}, \citenamefont {Terashima}, \citenamefont {Saito}, \citenamefont
  {Esparza~Echevarria}, \citenamefont {{Baptista de Castro}}, \citenamefont
  {Song}, \citenamefont {Iwasaki}, \citenamefont {Takeya},\ and\ \citenamefont
  {Takano}}]{adachiDemonstrationElectricDouble2020}%
  \BibitemOpen
  \bibfield  {author} {\bibinfo {author} {\bibfnamefont {S.}~\bibnamefont
  {Adachi}}, \bibinfo {author} {\bibfnamefont {R.}~\bibnamefont {Matsumoto}},
  \bibinfo {author} {\bibfnamefont {S.}~\bibnamefont {Yamamoto}}, \bibinfo
  {author} {\bibfnamefont {T.~D.}\ \bibnamefont {Yamamoto}}, \bibinfo {author}
  {\bibfnamefont {K.}~\bibnamefont {Terashima}}, \bibinfo {author}
  {\bibfnamefont {Y.}~\bibnamefont {Saito}}, \bibinfo {author} {\bibfnamefont
  {M.}~\bibnamefont {Esparza~Echevarria}}, \bibinfo {author} {\bibfnamefont
  {P.}~\bibnamefont {{Baptista de Castro}}}, \bibinfo {author} {\bibfnamefont
  {P.}~\bibnamefont {Song}}, \bibinfo {author} {\bibfnamefont {S.}~\bibnamefont
  {Iwasaki}}, \bibinfo {author} {\bibfnamefont {H.}~\bibnamefont {Takeya}}, \
  and\ \bibinfo {author} {\bibfnamefont {Y.}~\bibnamefont {Takano}},\ }\href
  {\doibase 10.1063/5.0004973} {\bibfield  {journal} {\bibinfo  {journal}
  {Applied Physics Letters}\ }\textbf {\bibinfo {volume} {116}},\ \bibinfo
  {pages} {223506} (\bibinfo {year} {2020})}\BibitemShut {NoStop}%
\bibitem [{\citenamefont {Matsumoto}\ \emph {et~al.}(2021)\citenamefont
  {Matsumoto}, \citenamefont {Nakano}, \citenamefont {Yamamoto},\ and\
  \citenamefont {Takano}}]{matsumotoSynthesisElectricalTransport2021}%
  \BibitemOpen
  \bibfield  {author} {\bibinfo {author} {\bibfnamefont {R.}~\bibnamefont
  {Matsumoto}}, \bibinfo {author} {\bibfnamefont {S.}~\bibnamefont {Nakano}},
  \bibinfo {author} {\bibfnamefont {S.}~\bibnamefont {Yamamoto}}, \ and\
  \bibinfo {author} {\bibfnamefont {Y.}~\bibnamefont {Takano}},\ }\href
  {\doibase 10.35848/1347-4065/ac1a49} {\bibfield  {journal} {\bibinfo
  {journal} {Jpn. J. Appl. Phys.}\ }\textbf {\bibinfo {volume} {60}},\ \bibinfo
  {pages} {090902} (\bibinfo {year} {2021})}\BibitemShut {NoStop}%
\bibitem [{\citenamefont {Mito}\ \emph {et~al.}(2001)\citenamefont {Mito},
  \citenamefont {Hitaka}, \citenamefont {Kawae}, \citenamefont {Takeda},
  \citenamefont {Kitai},\ and\ \citenamefont
  {Toyoshima}}]{mitoDevelopmentMiniatureDiamond2001}%
  \BibitemOpen
  \bibfield  {author} {\bibinfo {author} {\bibfnamefont {M.}~\bibnamefont
  {Mito}}, \bibinfo {author} {\bibfnamefont {M.}~\bibnamefont {Hitaka}},
  \bibinfo {author} {\bibfnamefont {T.}~\bibnamefont {Kawae}}, \bibinfo
  {author} {\bibfnamefont {K.}~\bibnamefont {Takeda}}, \bibinfo {author}
  {\bibfnamefont {T.}~\bibnamefont {Kitai}}, \ and\ \bibinfo {author}
  {\bibfnamefont {N.}~\bibnamefont {Toyoshima}},\ }\href {\doibase
  10.1143/JJAP.40.6641} {\bibfield  {journal} {\bibinfo  {journal} {Jpn. J.
  Appl. Phys.}\ }\textbf {\bibinfo {volume} {40}},\ \bibinfo {pages} {6641}
  (\bibinfo {year} {2001})}\BibitemShut {NoStop}%
\bibitem [{\citenamefont {Mito}\ \emph {et~al.}(2014)\citenamefont {Mito},
  \citenamefont {Imakyurei}, \citenamefont {Deguchi}, \citenamefont
  {Matsumoto}, \citenamefont {Hara}, \citenamefont {Ozaki}, \citenamefont
  {Takeya},\ and\ \citenamefont
  {Takano}}]{mitoEffectiveDisappearanceMeissner2014}%
  \BibitemOpen
  \bibfield  {author} {\bibinfo {author} {\bibfnamefont {M.}~\bibnamefont
  {Mito}}, \bibinfo {author} {\bibfnamefont {T.}~\bibnamefont {Imakyurei}},
  \bibinfo {author} {\bibfnamefont {H.}~\bibnamefont {Deguchi}}, \bibinfo
  {author} {\bibfnamefont {K.}~\bibnamefont {Matsumoto}}, \bibinfo {author}
  {\bibfnamefont {H.}~\bibnamefont {Hara}}, \bibinfo {author} {\bibfnamefont
  {T.}~\bibnamefont {Ozaki}}, \bibinfo {author} {\bibfnamefont
  {H.}~\bibnamefont {Takeya}}, \ and\ \bibinfo {author} {\bibfnamefont
  {Y.}~\bibnamefont {Takano}},\ }\href {\doibase 10.7566/JPSJ.83.023705}
  {\bibfield  {journal} {\bibinfo  {journal} {J. Phys. Soc. Jpn.}\ }\textbf
  {\bibinfo {volume} {83}},\ \bibinfo {pages} {023705} (\bibinfo {year}
  {2014})}\BibitemShut {NoStop}%
\bibitem [{\citenamefont {Mito}\ \emph {et~al.}(2016)\citenamefont {Mito},
  \citenamefont {Goto}, \citenamefont {Matsui}, \citenamefont {Deguchi},
  \citenamefont {Matsumoto}, \citenamefont {Hara}, \citenamefont {Ozaki},
  \citenamefont {Takeya},\ and\ \citenamefont
  {Takano}}]{mitoUniaxialStrainEffects2016}%
  \BibitemOpen
  \bibfield  {author} {\bibinfo {author} {\bibfnamefont {M.}~\bibnamefont
  {Mito}}, \bibinfo {author} {\bibfnamefont {H.}~\bibnamefont {Goto}}, \bibinfo
  {author} {\bibfnamefont {H.}~\bibnamefont {Matsui}}, \bibinfo {author}
  {\bibfnamefont {H.}~\bibnamefont {Deguchi}}, \bibinfo {author} {\bibfnamefont
  {K.}~\bibnamefont {Matsumoto}}, \bibinfo {author} {\bibfnamefont
  {H.}~\bibnamefont {Hara}}, \bibinfo {author} {\bibfnamefont {T.}~\bibnamefont
  {Ozaki}}, \bibinfo {author} {\bibfnamefont {H.}~\bibnamefont {Takeya}}, \
  and\ \bibinfo {author} {\bibfnamefont {Y.}~\bibnamefont {Takano}},\ }\href
  {\doibase 10.7566/JPSJ.85.024711} {\bibfield  {journal} {\bibinfo  {journal}
  {J. Phys. Soc. Jpn.}\ }\textbf {\bibinfo {volume} {85}},\ \bibinfo {pages}
  {024711} (\bibinfo {year} {2016})}\BibitemShut {NoStop}%
\bibitem [{\citenamefont {Mito}\ \emph {et~al.}(2017)\citenamefont {Mito},
  \citenamefont {Ogata}, \citenamefont {Goto}, \citenamefont {Tsuruta},
  \citenamefont {Nakamura}, \citenamefont {Deguchi}, \citenamefont {Horide},
  \citenamefont {Matsumoto}, \citenamefont {Tajiri}, \citenamefont {Hara},
  \citenamefont {Ozaki}, \citenamefont {Takeya},\ and\ \citenamefont
  {Takano}}]{mitoUniaxialStrainEffects2017}%
  \BibitemOpen
  \bibfield  {author} {\bibinfo {author} {\bibfnamefont {M.}~\bibnamefont
  {Mito}}, \bibinfo {author} {\bibfnamefont {K.}~\bibnamefont {Ogata}},
  \bibinfo {author} {\bibfnamefont {H.}~\bibnamefont {Goto}}, \bibinfo {author}
  {\bibfnamefont {K.}~\bibnamefont {Tsuruta}}, \bibinfo {author} {\bibfnamefont
  {K.}~\bibnamefont {Nakamura}}, \bibinfo {author} {\bibfnamefont
  {H.}~\bibnamefont {Deguchi}}, \bibinfo {author} {\bibfnamefont
  {T.}~\bibnamefont {Horide}}, \bibinfo {author} {\bibfnamefont
  {K.}~\bibnamefont {Matsumoto}}, \bibinfo {author} {\bibfnamefont
  {T.}~\bibnamefont {Tajiri}}, \bibinfo {author} {\bibfnamefont
  {H.}~\bibnamefont {Hara}}, \bibinfo {author} {\bibfnamefont {T.}~\bibnamefont
  {Ozaki}}, \bibinfo {author} {\bibfnamefont {H.}~\bibnamefont {Takeya}}, \
  and\ \bibinfo {author} {\bibfnamefont {Y.}~\bibnamefont {Takano}},\ }\href
  {\doibase 10.1103/PhysRevB.95.064503} {\bibfield  {journal} {\bibinfo
  {journal} {Phys. Rev. B}\ }\textbf {\bibinfo {volume} {95}},\ \bibinfo
  {pages} {064503} (\bibinfo {year} {2017})}\BibitemShut {NoStop}%
\bibitem [{\citenamefont {{Abdel-Hafiez}}\ \emph {et~al.}(2018)\citenamefont
  {{Abdel-Hafiez}}, \citenamefont {Zhao}, \citenamefont {Huang}, \citenamefont
  {Cho}, \citenamefont {Wong}, \citenamefont {Hassen}, \citenamefont {Ohkuma},
  \citenamefont {Fang}, \citenamefont {Pan}, \citenamefont {Ren}, \citenamefont
  {Sadakov}, \citenamefont {Usoltsev}, \citenamefont {Pudalov}, \citenamefont
  {Mito}, \citenamefont {Lortz}, \citenamefont {Krellner},\ and\ \citenamefont
  {Yang}}]{abdel-hafiezHighpressureEffectsIsotropic2018}%
  \BibitemOpen
  \bibfield  {author} {\bibinfo {author} {\bibfnamefont {M.}~\bibnamefont
  {{Abdel-Hafiez}}}, \bibinfo {author} {\bibfnamefont {Y.}~\bibnamefont
  {Zhao}}, \bibinfo {author} {\bibfnamefont {Z.}~\bibnamefont {Huang}},
  \bibinfo {author} {\bibfnamefont {C.-w.}\ \bibnamefont {Cho}}, \bibinfo
  {author} {\bibfnamefont {C.~H.}\ \bibnamefont {Wong}}, \bibinfo {author}
  {\bibfnamefont {A.}~\bibnamefont {Hassen}}, \bibinfo {author} {\bibfnamefont
  {M.}~\bibnamefont {Ohkuma}}, \bibinfo {author} {\bibfnamefont {Y.-W.}\
  \bibnamefont {Fang}}, \bibinfo {author} {\bibfnamefont {B.-J.}\ \bibnamefont
  {Pan}}, \bibinfo {author} {\bibfnamefont {Z.-A.}\ \bibnamefont {Ren}},
  \bibinfo {author} {\bibfnamefont {A.}~\bibnamefont {Sadakov}}, \bibinfo
  {author} {\bibfnamefont {A.}~\bibnamefont {Usoltsev}}, \bibinfo {author}
  {\bibfnamefont {V.}~\bibnamefont {Pudalov}}, \bibinfo {author} {\bibfnamefont
  {M.}~\bibnamefont {Mito}}, \bibinfo {author} {\bibfnamefont {R.}~\bibnamefont
  {Lortz}}, \bibinfo {author} {\bibfnamefont {C.}~\bibnamefont {Krellner}}, \
  and\ \bibinfo {author} {\bibfnamefont {W.}~\bibnamefont {Yang}},\ }\href
  {\doibase 10.1103/PhysRevB.97.134508} {\bibfield  {journal} {\bibinfo
  {journal} {Phys. Rev. B}\ }\textbf {\bibinfo {volume} {97}},\ \bibinfo
  {pages} {134508} (\bibinfo {year} {2018})}\BibitemShut {NoStop}%
\bibitem [{\citenamefont {Mito}\ \emph {et~al.}(2019)\citenamefont {Mito},
  \citenamefont {Kitamura}, \citenamefont {Tajiri}, \citenamefont {Nakamura},
  \citenamefont {Shiraishi}, \citenamefont {Ogata}, \citenamefont {Deguchi},
  \citenamefont {Yamaguchi}, \citenamefont {Takeshita}, \citenamefont
  {Nishizaki}, \citenamefont {Edalati},\ and\ \citenamefont
  {Horita}}]{mitoHydrostaticPressureEffects2019}%
  \BibitemOpen
  \bibfield  {author} {\bibinfo {author} {\bibfnamefont {M.}~\bibnamefont
  {Mito}}, \bibinfo {author} {\bibfnamefont {Y.}~\bibnamefont {Kitamura}},
  \bibinfo {author} {\bibfnamefont {T.}~\bibnamefont {Tajiri}}, \bibinfo
  {author} {\bibfnamefont {K.}~\bibnamefont {Nakamura}}, \bibinfo {author}
  {\bibfnamefont {R.}~\bibnamefont {Shiraishi}}, \bibinfo {author}
  {\bibfnamefont {K.}~\bibnamefont {Ogata}}, \bibinfo {author} {\bibfnamefont
  {H.}~\bibnamefont {Deguchi}}, \bibinfo {author} {\bibfnamefont
  {T.}~\bibnamefont {Yamaguchi}}, \bibinfo {author} {\bibfnamefont
  {N.}~\bibnamefont {Takeshita}}, \bibinfo {author} {\bibfnamefont
  {T.}~\bibnamefont {Nishizaki}}, \bibinfo {author} {\bibfnamefont
  {K.}~\bibnamefont {Edalati}}, \ and\ \bibinfo {author} {\bibfnamefont
  {Z.}~\bibnamefont {Horita}},\ }\href {\doibase 10.1063/1.5083094} {\bibfield
  {journal} {\bibinfo  {journal} {Journal of Applied Physics}\ }\textbf
  {\bibinfo {volume} {125}},\ \bibinfo {pages} {125901} (\bibinfo {year}
  {2019})}\BibitemShut {NoStop}%
\bibitem [{\citenamefont {Irifune}\ \emph {et~al.}(2003)\citenamefont
  {Irifune}, \citenamefont {Kurio}, \citenamefont {Sakamoto}, \citenamefont
  {Inoue},\ and\ \citenamefont
  {Sumiya}}]{irifuneUltrahardPolycrystallineDiamond2003}%
  \BibitemOpen
  \bibfield  {author} {\bibinfo {author} {\bibfnamefont {T.}~\bibnamefont
  {Irifune}}, \bibinfo {author} {\bibfnamefont {A.}~\bibnamefont {Kurio}},
  \bibinfo {author} {\bibfnamefont {S.}~\bibnamefont {Sakamoto}}, \bibinfo
  {author} {\bibfnamefont {T.}~\bibnamefont {Inoue}}, \ and\ \bibinfo {author}
  {\bibfnamefont {H.}~\bibnamefont {Sumiya}},\ }\href {\doibase
  10.1038/421599b} {\bibfield  {journal} {\bibinfo  {journal} {Nature}\
  }\textbf {\bibinfo {volume} {421}},\ \bibinfo {pages} {599} (\bibinfo {year}
  {2003})}\BibitemShut {NoStop}%
\bibitem [{\citenamefont {Dolan}\ and\ \citenamefont
  {Silcox}(1973)}]{dolanCriticalThicknessesSuperconducting1973}%
  \BibitemOpen
  \bibfield  {author} {\bibinfo {author} {\bibfnamefont {G.~J.}\ \bibnamefont
  {Dolan}}\ and\ \bibinfo {author} {\bibfnamefont {J.}~\bibnamefont {Silcox}},\
  }\href {\doibase 10.1103/PhysRevLett.30.603} {\bibfield  {journal} {\bibinfo
  {journal} {Phys. Rev. Lett.}\ }\textbf {\bibinfo {volume} {30}},\ \bibinfo
  {pages} {603} (\bibinfo {year} {1973})}\BibitemShut {NoStop}%
\bibitem [{\citenamefont {Ohkuma}\ \emph {et~al.}(2023)\citenamefont {Ohkuma},
  \citenamefont {Matsumoto},\ and\ \citenamefont
  {Takano}}]{ohkumaNonreciprocalSupercurrentThin2023}%
  \BibitemOpen
  \bibfield  {author} {\bibinfo {author} {\bibfnamefont {M.}~\bibnamefont
  {Ohkuma}}, \bibinfo {author} {\bibfnamefont {R.}~\bibnamefont {Matsumoto}}, \
  and\ \bibinfo {author} {\bibfnamefont {Y.}~\bibnamefont {Takano}},\ }\href
  {\doibase 10.35848/1882-0786/acc8b5} {\bibfield  {journal} {\bibinfo
  {journal} {Appl. Phys. Express}\ }\textbf {\bibinfo {volume} {16}},\ \bibinfo
  {pages} {043004} (\bibinfo {year} {2023})}\BibitemShut {NoStop}%
\bibitem [{\citenamefont {Helfand}\ and\ \citenamefont
  {Werthamer}(1966)}]{helfandTemperaturePurityDependence1966}%
  \BibitemOpen
  \bibfield  {author} {\bibinfo {author} {\bibfnamefont {E.}~\bibnamefont
  {Helfand}}\ and\ \bibinfo {author} {\bibfnamefont {N.~R.}\ \bibnamefont
  {Werthamer}},\ }\href {\doibase 10.1103/PhysRev.147.288} {\bibfield
  {journal} {\bibinfo  {journal} {Phys. Rev.}\ }\textbf {\bibinfo {volume}
  {147}},\ \bibinfo {pages} {288} (\bibinfo {year} {1966})}\BibitemShut
  {NoStop}%
\bibitem [{\citenamefont {Werthamer}\ \emph {et~al.}(1966)\citenamefont
  {Werthamer}, \citenamefont {Helfand},\ and\ \citenamefont
  {Hohenberg}}]{werthamerTemperaturePurityDependence1966}%
  \BibitemOpen
  \bibfield  {author} {\bibinfo {author} {\bibfnamefont {N.~R.}\ \bibnamefont
  {Werthamer}}, \bibinfo {author} {\bibfnamefont {E.}~\bibnamefont {Helfand}},
  \ and\ \bibinfo {author} {\bibfnamefont {P.~C.}\ \bibnamefont {Hohenberg}},\
  }\href {\doibase 10.1103/PhysRev.147.295} {\bibfield  {journal} {\bibinfo
  {journal} {Phys. Rev.}\ }\textbf {\bibinfo {volume} {147}},\ \bibinfo {pages}
  {295} (\bibinfo {year} {1966})}\BibitemShut {NoStop}%
\bibitem [{\citenamefont {Baumgartner}\ \emph {et~al.}(2014)\citenamefont
  {Baumgartner}, \citenamefont {Eisterer}, \citenamefont {Weber}, \citenamefont
  {Fl{\"u}kiger}, \citenamefont {Scheuerlein},\ and\ \citenamefont
  {Bottura}}]{baumgartnerEffectsNeutronIrradiation2013}%
  \BibitemOpen
  \bibfield  {author} {\bibinfo {author} {\bibfnamefont {T.}~\bibnamefont
  {Baumgartner}}, \bibinfo {author} {\bibfnamefont {M.}~\bibnamefont
  {Eisterer}}, \bibinfo {author} {\bibfnamefont {H.~W.}\ \bibnamefont {Weber}},
  \bibinfo {author} {\bibfnamefont {R.}~\bibnamefont {Fl{\"u}kiger}}, \bibinfo
  {author} {\bibfnamefont {C.}~\bibnamefont {Scheuerlein}}, \ and\ \bibinfo
  {author} {\bibfnamefont {L.}~\bibnamefont {Bottura}},\ }\href {\doibase
  10.1088/0953-2048/27/1/015005} {\bibfield  {journal} {\bibinfo  {journal}
  {Supercond. Sci. Technol.}\ }\textbf {\bibinfo {volume} {27}},\ \bibinfo
  {pages} {015005} (\bibinfo {year} {2014})}\BibitemShut {NoStop}%
\bibitem [{\citenamefont {Zhang}\ \emph {et~al.}(2016)\citenamefont {Zhang},
  \citenamefont {Wong}, \citenamefont {Shen}, \citenamefont {Sze},
  \citenamefont {Zhang}, \citenamefont {Zhang}, \citenamefont {Dong},
  \citenamefont {Xu}, \citenamefont {Yan}, \citenamefont {Li}, \citenamefont
  {Hu},\ and\ \citenamefont
  {Lortz}}]{zhangDramaticEnhancementSuperconductivity2016}%
  \BibitemOpen
  \bibfield  {author} {\bibinfo {author} {\bibfnamefont {Y.}~\bibnamefont
  {Zhang}}, \bibinfo {author} {\bibfnamefont {C.~H.}\ \bibnamefont {Wong}},
  \bibinfo {author} {\bibfnamefont {J.}~\bibnamefont {Shen}}, \bibinfo {author}
  {\bibfnamefont {S.~T.}\ \bibnamefont {Sze}}, \bibinfo {author} {\bibfnamefont
  {B.}~\bibnamefont {Zhang}}, \bibinfo {author} {\bibfnamefont
  {H.}~\bibnamefont {Zhang}}, \bibinfo {author} {\bibfnamefont
  {Y.}~\bibnamefont {Dong}}, \bibinfo {author} {\bibfnamefont {H.}~\bibnamefont
  {Xu}}, \bibinfo {author} {\bibfnamefont {Z.}~\bibnamefont {Yan}}, \bibinfo
  {author} {\bibfnamefont {Y.}~\bibnamefont {Li}}, \bibinfo {author}
  {\bibfnamefont {X.}~\bibnamefont {Hu}}, \ and\ \bibinfo {author}
  {\bibfnamefont {R.}~\bibnamefont {Lortz}},\ }\href {\doibase
  10.1038/srep32963} {\bibfield  {journal} {\bibinfo  {journal} {Sci Rep}\
  }\textbf {\bibinfo {volume} {6}},\ \bibinfo {pages} {32963} (\bibinfo {year}
  {2016})}\BibitemShut {NoStop}%
\bibitem [{\citenamefont {Shao}\ \emph {et~al.}(2018)\citenamefont {Shao},
  \citenamefont {Zhang}, \citenamefont {Yuan}, \citenamefont {Sun},
  \citenamefont {Cao}, \citenamefont {Zhang}, \citenamefont {Li}, \citenamefont
  {Gedeon}, \citenamefont {Xiang}, \citenamefont {Xue},\ and\ \citenamefont
  {Pan}}]{shaoScanningTunnelingMicroscopic2018a}%
  \BibitemOpen
  \bibfield  {author} {\bibinfo {author} {\bibfnamefont {Z.}~\bibnamefont
  {Shao}}, \bibinfo {author} {\bibfnamefont {Z.}~\bibnamefont {Zhang}},
  \bibinfo {author} {\bibfnamefont {H.}~\bibnamefont {Yuan}}, \bibinfo {author}
  {\bibfnamefont {H.}~\bibnamefont {Sun}}, \bibinfo {author} {\bibfnamefont
  {Y.}~\bibnamefont {Cao}}, \bibinfo {author} {\bibfnamefont {X.}~\bibnamefont
  {Zhang}}, \bibinfo {author} {\bibfnamefont {S.}~\bibnamefont {Li}}, \bibinfo
  {author} {\bibfnamefont {H.}~\bibnamefont {Gedeon}}, \bibinfo {author}
  {\bibfnamefont {T.}~\bibnamefont {Xiang}}, \bibinfo {author} {\bibfnamefont
  {Q.-K.}\ \bibnamefont {Xue}}, \ and\ \bibinfo {author} {\bibfnamefont
  {M.}~\bibnamefont {Pan}},\ }\href {\doibase 10.1016/j.scib.2018.09.006}
  {\bibfield  {journal} {\bibinfo  {journal} {Science Bulletin}\ }\textbf
  {\bibinfo {volume} {63}},\ \bibinfo {pages} {1332} (\bibinfo {year}
  {2018})}\BibitemShut {NoStop}%
\bibitem [{\citenamefont {Glover}(1967)}]{gloverIdealResistiveTransition1967}%
  \BibitemOpen
  \bibfield  {author} {\bibinfo {author} {\bibfnamefont {R.~E.}\ \bibnamefont
  {Glover}},\ }\href {\doibase 10.1016/0375-9601(67)90036-9} {\bibfield
  {journal} {\bibinfo  {journal} {Physics Letters A}\ }\textbf {\bibinfo
  {volume} {25}},\ \bibinfo {pages} {542} (\bibinfo {year} {1967})}\BibitemShut
  {NoStop}%
\bibitem [{\citenamefont {Aslamasov}\ and\ \citenamefont
  {Larkin}(1968)}]{aslamasovInfluenceFluctuationPairing1968}%
  \BibitemOpen
  \bibfield  {author} {\bibinfo {author} {\bibfnamefont {L.~G.}\ \bibnamefont
  {Aslamasov}}\ and\ \bibinfo {author} {\bibfnamefont {A.~I.}\ \bibnamefont
  {Larkin}},\ }\href {\doibase 10.1016/0375-9601(68)90623-3} {\bibfield
  {journal} {\bibinfo  {journal} {Physics Letters A}\ }\textbf {\bibinfo
  {volume} {26}},\ \bibinfo {pages} {238} (\bibinfo {year} {1968})}\BibitemShut
  {NoStop}%
\bibitem [{\citenamefont {Skocpol}\ and\ \citenamefont
  {Tinkham}(1975)}]{skocpolFluctuationsSuperconductingPhase1975}%
  \BibitemOpen
  \bibfield  {author} {\bibinfo {author} {\bibfnamefont {W.~J.}\ \bibnamefont
  {Skocpol}}\ and\ \bibinfo {author} {\bibfnamefont {M.}~\bibnamefont
  {Tinkham}},\ }\href {\doibase 10.1088/0034-4885/38/9/001} {\bibfield
  {journal} {\bibinfo  {journal} {Rep. Prog. Phys.}\ }\textbf {\bibinfo
  {volume} {38}},\ \bibinfo {pages} {1049} (\bibinfo {year}
  {1975})}\BibitemShut {NoStop}%
\bibitem [{\citenamefont {S{\'o}{\~n}ora}\ \emph {et~al.}(2019)\citenamefont
  {S{\'o}{\~n}ora}, \citenamefont {Carballeira}, \citenamefont {Ponte},
  \citenamefont {Vidal}, \citenamefont {Grenet},\ and\ \citenamefont
  {Mosqueira}}]{sonoraParaconductivityGranularFilms2019}%
  \BibitemOpen
  \bibfield  {author} {\bibinfo {author} {\bibfnamefont {D.}~\bibnamefont
  {S{\'o}{\~n}ora}}, \bibinfo {author} {\bibfnamefont {C.}~\bibnamefont
  {Carballeira}}, \bibinfo {author} {\bibfnamefont {J.~J.}\ \bibnamefont
  {Ponte}}, \bibinfo {author} {\bibfnamefont {F.}~\bibnamefont {Vidal}},
  \bibinfo {author} {\bibfnamefont {T.}~\bibnamefont {Grenet}}, \ and\ \bibinfo
  {author} {\bibfnamefont {J.}~\bibnamefont {Mosqueira}},\ }\href {\doibase
  10.1103/PhysRevB.100.104509} {\bibfield  {journal} {\bibinfo  {journal}
  {Phys. Rev. B}\ }\textbf {\bibinfo {volume} {100}},\ \bibinfo {pages}
  {104509} (\bibinfo {year} {2019})}\BibitemShut {NoStop}%
\bibitem [{\citenamefont {Hsieh}\ \emph {et~al.}(2019)\citenamefont {Hsieh},
  \citenamefont {Bhattacharyya}, \citenamefont {Zu}, \citenamefont {Mittiga},
  \citenamefont {Smart}, \citenamefont {Machado}, \citenamefont {Kobrin},
  \citenamefont {H{\"o}hn}, \citenamefont {Rui}, \citenamefont {Kamrani},
  \citenamefont {Chatterjee}, \citenamefont {Choi}, \citenamefont {Zaletel},
  \citenamefont {Struzhkin}, \citenamefont {Moore}, \citenamefont {Levitas},
  \citenamefont {Jeanloz},\ and\ \citenamefont
  {Yao}}]{hsiehImagingStressMagnetism2019}%
  \BibitemOpen
  \bibfield  {author} {\bibinfo {author} {\bibfnamefont {S.}~\bibnamefont
  {Hsieh}}, \bibinfo {author} {\bibfnamefont {P.}~\bibnamefont
  {Bhattacharyya}}, \bibinfo {author} {\bibfnamefont {C.}~\bibnamefont {Zu}},
  \bibinfo {author} {\bibfnamefont {T.}~\bibnamefont {Mittiga}}, \bibinfo
  {author} {\bibfnamefont {T.~J.}\ \bibnamefont {Smart}}, \bibinfo {author}
  {\bibfnamefont {F.}~\bibnamefont {Machado}}, \bibinfo {author} {\bibfnamefont
  {B.}~\bibnamefont {Kobrin}}, \bibinfo {author} {\bibfnamefont {T.~O.}\
  \bibnamefont {H{\"o}hn}}, \bibinfo {author} {\bibfnamefont {N.~Z.}\
  \bibnamefont {Rui}}, \bibinfo {author} {\bibfnamefont {M.}~\bibnamefont
  {Kamrani}}, \bibinfo {author} {\bibfnamefont {S.}~\bibnamefont {Chatterjee}},
  \bibinfo {author} {\bibfnamefont {S.}~\bibnamefont {Choi}}, \bibinfo {author}
  {\bibfnamefont {M.}~\bibnamefont {Zaletel}}, \bibinfo {author} {\bibfnamefont
  {V.~V.}\ \bibnamefont {Struzhkin}}, \bibinfo {author} {\bibfnamefont {J.~E.}\
  \bibnamefont {Moore}}, \bibinfo {author} {\bibfnamefont {V.~I.}\ \bibnamefont
  {Levitas}}, \bibinfo {author} {\bibfnamefont {R.}~\bibnamefont {Jeanloz}}, \
  and\ \bibinfo {author} {\bibfnamefont {N.~Y.}\ \bibnamefont {Yao}},\ }\href
  {\doibase 10.1126/science.aaw4352} {\bibfield  {journal} {\bibinfo  {journal}
  {Science}\ }\textbf {\bibinfo {volume} {366}},\ \bibinfo {pages} {1349}
  (\bibinfo {year} {2019})}\BibitemShut {NoStop}%
\bibitem [{\citenamefont {Lesik}\ \emph {et~al.}(2019)\citenamefont {Lesik},
  \citenamefont {Plisson}, \citenamefont {Toraille}, \citenamefont {Renaud},
  \citenamefont {Occelli}, \citenamefont {Schmidt}, \citenamefont {Salord},
  \citenamefont {Delobbe}, \citenamefont {Debuisschert}, \citenamefont
  {Rondin}, \citenamefont {Loubeyre},\ and\ \citenamefont
  {Roch}}]{lesikMagneticMeasurementsMicrometersized2019}%
  \BibitemOpen
  \bibfield  {author} {\bibinfo {author} {\bibfnamefont {M.}~\bibnamefont
  {Lesik}}, \bibinfo {author} {\bibfnamefont {T.}~\bibnamefont {Plisson}},
  \bibinfo {author} {\bibfnamefont {L.}~\bibnamefont {Toraille}}, \bibinfo
  {author} {\bibfnamefont {J.}~\bibnamefont {Renaud}}, \bibinfo {author}
  {\bibfnamefont {F.}~\bibnamefont {Occelli}}, \bibinfo {author} {\bibfnamefont
  {M.}~\bibnamefont {Schmidt}}, \bibinfo {author} {\bibfnamefont
  {O.}~\bibnamefont {Salord}}, \bibinfo {author} {\bibfnamefont
  {A.}~\bibnamefont {Delobbe}}, \bibinfo {author} {\bibfnamefont
  {T.}~\bibnamefont {Debuisschert}}, \bibinfo {author} {\bibfnamefont
  {L.}~\bibnamefont {Rondin}}, \bibinfo {author} {\bibfnamefont
  {P.}~\bibnamefont {Loubeyre}}, \ and\ \bibinfo {author} {\bibfnamefont
  {J.-F.}\ \bibnamefont {Roch}},\ }\href {\doibase 10.1126/science.aaw4329}
  {\bibfield  {journal} {\bibinfo  {journal} {Science}\ }\textbf {\bibinfo
  {volume} {366}},\ \bibinfo {pages} {1359} (\bibinfo {year}
  {2019})}\BibitemShut {NoStop}%
\bibitem [{\citenamefont {Umeo}(2016)}]{umeoAlternatingCurrentCalorimeter2016}%
  \BibitemOpen
  \bibfield  {author} {\bibinfo {author} {\bibfnamefont {K.}~\bibnamefont
  {Umeo}},\ }\href {\doibase 10.1063/1.4952959} {\bibfield  {journal} {\bibinfo
   {journal} {Review of Scientific Instruments}\ }\textbf {\bibinfo {volume}
  {87}},\ \bibinfo {pages} {063901} (\bibinfo {year} {2016})}\BibitemShut
  {NoStop}%
\bibitem [{\citenamefont {Cao}\ \emph {et~al.}(2023)\citenamefont {Cao},
  \citenamefont {Jang}, \citenamefont {Choi}, \citenamefont {Kim},
  \citenamefont {Kim}, \citenamefont {Zhang}, \citenamefont {Sharbirin},
  \citenamefont {Kim},\ and\ \citenamefont
  {Park}}]{caoSpectroscopicEvidenceSuperconductivity2023}%
  \BibitemOpen
  \bibfield  {author} {\bibinfo {author} {\bibfnamefont {Z.-Y.}\ \bibnamefont
  {Cao}}, \bibinfo {author} {\bibfnamefont {H.}~\bibnamefont {Jang}}, \bibinfo
  {author} {\bibfnamefont {S.}~\bibnamefont {Choi}}, \bibinfo {author}
  {\bibfnamefont {J.}~\bibnamefont {Kim}}, \bibinfo {author} {\bibfnamefont
  {S.}~\bibnamefont {Kim}}, \bibinfo {author} {\bibfnamefont {J.-B.}\
  \bibnamefont {Zhang}}, \bibinfo {author} {\bibfnamefont {A.~S.}\ \bibnamefont
  {Sharbirin}}, \bibinfo {author} {\bibfnamefont {J.}~\bibnamefont {Kim}}, \
  and\ \bibinfo {author} {\bibfnamefont {T.}~\bibnamefont {Park}},\ }\href
  {\doibase 10.1038/s41427-022-00457-6} {\bibfield  {journal} {\bibinfo
  {journal} {NPG Asia Mater}\ }\textbf {\bibinfo {volume} {15}},\ \bibinfo
  {pages} {1} (\bibinfo {year} {2023})}\BibitemShut {NoStop}%
\bibitem [{\citenamefont {Sasaki}\ \emph {et~al.}(2025)\citenamefont {Sasaki},
  \citenamefont {Ohkuma}, \citenamefont {Matsumoto}, \citenamefont {Shinmei},
  \citenamefont {Irifune}, \citenamefont {Takano},\ and\ \citenamefont
  {Shimizu}}]{sasakiEnhancementSuperconductivityThin2025data}%
  \BibitemOpen
  \bibfield  {author} {\bibinfo {author} {\bibfnamefont {M.}~\bibnamefont
  {Sasaki}}, \bibinfo {author} {\bibfnamefont {M.}~\bibnamefont {Ohkuma}},
  \bibinfo {author} {\bibfnamefont {R.}~\bibnamefont {Matsumoto}}, \bibinfo
  {author} {\bibfnamefont {T.}~\bibnamefont {Shinmei}}, \bibinfo {author}
  {\bibfnamefont {T.}~\bibnamefont {Irifune}}, \bibinfo {author} {\bibfnamefont
  {Y.}~\bibnamefont {Takano}}, \ and\ \bibinfo {author} {\bibfnamefont
  {K.}~\bibnamefont {Shimizu}},\ }\href {\doibase
  https://zenodo.org/records/14759622} {\  (\bibinfo {year} {2025}),\
  https://zenodo.org/records/14759622}\BibitemShut {NoStop}%
\end{thebibliography}%

%
%

\end{document}